\documentclass[aps,floatfix,twocolumn,showpacs,pra]{revtex4}
\usepackage{psfrag}
\usepackage{graphicx,lipsum} %
\usepackage{amsmath,color}
\usepackage{pgfplots} %

\usepackage[T1]{fontenc} %
\usepackage[latin2]{inputenc} %

\usepackage[normalem]{ulem}

\def\gappeq{\mathrel{ \rlap{\raise.5ex\hbox{$>$}}
		{\lower.5ex\hbox{$\sim$}} } }
\def\lappeq{\mathrel{ \rlap{\raise.5ex\hbox{$<$}}
		{\lower.5ex\hbox{$\sim$}} } }
\usepackage{bm}
\usepackage{color}
\usepackage{epstopdf}

\usepackage{amsmath}%
\usepackage{amsfonts}%
\usepackage{amssymb}%
\usepackage{graphicx}
\usepackage{array}%

\usepackage{epsfig}
\usepackage{units}
\usepackage{url}
\usepackage[breaklinks=true,colorlinks=true,linkcolor=blue,urlcolor=blue,citecolor=blue]{hyperref}

\usepackage{bm}

\newcommand{\be}{\begin{equation}}
\newcommand{\ee}{\end{equation}}
\newcommand{\bea}{\begin{eqnarray}}
\newcommand{\eea}{\end{eqnarray}}
\newcommand{\rr}{\mathbf{r}}
\newcommand{\kk}{\mathbf{k}}
\newcommand{\qq}{\mathbf{q}}

\newcommand*{\Sz}{\ensuremath{\hat{S}_z}}

\newcommand{\bra}[1]{\left\langle#1\right|}  	
\newcommand{\ket}[1]{\left|#1\right\rangle}   	
\newcommand{\bb}[1]{\left(#1\right)}

\newcommand{\braket}[2]{\left\langle#1\middle|#2\right\rangle}

\newcommand{\gt}{\tilde{\gamma}}                                    

\newcommand{\tcat}{t_{{\rm cat}}}
\newcommand{\ketph}[2]{\ket{#1 ;\, #2}}

\newcommand{\OO}{\mathbf{0}}

\def\rien{\color{red}}

\def\rien{\color{black}}

\begin{document}
\title{Mesoscopic {\rien quantum} superpositions in bimodal Bose-Einstein condensates: {\rien decoherence and} strategies to counteract {\rien it}}
\author{K. Paw\l{}owski}
\affiliation{Center for Theoretical Physics, Polish Academy of Sciences, Al. Lotnik\'ow 32/46, 02-668 Warsaw, Poland}
\author{Matteo Fadel}
\affiliation{Department of Physics, University of Basel, Klingelbergstrasse 82, CH-4056 Basel}
\author{Philipp Treutlein}
\affiliation{Department of Physics, University of Basel, Klingelbergstrasse 82, CH-4056 Basel}
\author{Y. Castin}
\affiliation{Laboratoire Kastler Brossel, ENS-PSL, CNRS, UPMC-Sorbonne Universit\'es and Coll\`ege de France, Paris, France}
\author{A. Sinatra}
\affiliation{Laboratoire Kastler Brossel, ENS-PSL, CNRS, UPMC-Sorbonne Universit\'es and Coll\`ege de France, Paris, France}

\begin{abstract}
We study {\rien theoretically} the interaction-induced generation of {\rien mesoscopic coherent spin state superpositions (small-particle-number cat states) from an initial  coherent spin state}
in bimodal Bose-Einstein condensates {\rien and the subsequent phase revival}, 
including decoherence {\rien due to}
particle losses and fluctuations of the total {\rien particle number}.
{\rien In a full multimode description, we propose a preparation procedure of the initial
{\rien coherent spin state} and we study the effect of preexisting thermal fluctuations {\rien on the phase revival, and  on the spin and orbito-spinorial cat-state fidelities}.}
\end{abstract}

\maketitle


\section{Introduction}
{\rien While mesoscopic superpositions of coherent states of light with up to one hundred photons \cite{Haroche,Schoelkopf,Vlastakis},
and Greenberger-Horne-Zeilinger states with up to 14 trapped ions \cite{Leibfried,Monz} have been generated and observed in experiments, Schr\"odinger cat states with atomic gases are still out of reach \cite{Rev_Pezze}}.
Bose-Einstein condensates of ultra-cold atoms, confined in {\rien conservative potentials made by light or magnetic fields,} are excellent candidates to take up the challenge as they are to a good approximation isolated systems. 
{\rien Characterized by a macroscopic population of a single particle state,} condensates offer in principle the unprecedented possibility of generating large  {\it orbito-spinorial} Schr\"odiger cats, that is superpositions of two coherent spin states with opposite phases, each in a single and well controlled quantum state concerning the orbital degrees of freedom.
Nevertheless, decoherence
originating from particle losses and total atom number fluctuations, as well as from the intrinsic multimode nature of the atomic field and {\rien nonzero initial} temperature, is {\rien usually not} negligible.
The aim of this work is to present strategies to counteract decoherence, within the possibilities and constraints of specific experiments on bimodal condensates. {\rien Other proposals starting with monomode condensates,
see e.g.\, \cite{Weiss}, are not discussed here.}

In analogy to the well-known optical proposal of {\rien Yurke and Stoler of 1986 in reference \cite{Yurke}}, in bimodal condensates
the entanglement  {\rien stems} from the interactions between atoms that introduce a nonlinearity of the Kerr type for the atomic field. A state where all the atoms are in a superposition of the two modes with a well defined relative phase, a so-called ``phase state'' {\rien or ``coherent spin state''}, dynamically evolves into a Schr\"odinger cat state, superposition of two phase states with opposite relative phases {\rien \cite{Moelmer,LesHouches}}. 
At twice the cat-state time the system {\rien returns} into a single phase state giving rise to a revival peak in the contrast of the interference pattern between the two modes \cite{DalibardCastin}.

The influence of particle losses on the revival peak amplitude has been studied analytically in reference \cite{Sinatra98}.
In section \ref{sec:CLRA} of the present paper we show that there is a simple quantitative relation between the amplitude of the revival peak and the cat-state fidelity. As an application, for $N=300$ {\rien rubidium 87 atoms} in two separated spatial modes, {\rien with three-body losses} but no fluctuations of the total number of particles and at zero temperature, we calculate that a Schr\"odinger cat is obtained {\rien in time $t_{\rm cat}=128$ ms} with a fidelity 
{\rien ${\cal F} \,{\rien \simeq} \, 0.8$}.

In section {\rien \ref{sec:HTWROSA}} we concentrate on {\rien the use of} two internal states of a hyperfine transition in sodium or rubidium atoms. For Rb we consider the states $|F=1,m_F=-1\rangle$ and $|F=2,m_F=1\rangle$ that have been used to generate spin squeezing in state-dependent potentials on a chip \cite{SqMunich,MWpotentials}. A particular interest in these states resides in the fact that they {\rien form the clock transition}
in atomic clock experiments with trapped atoms on a chip \cite{TACC}. We {\rien present} a strategy to obtain mesoscopic superpositions using these systems, despite severe {\rien intrinsic} and experimental constraints, including particle losses and Poissonian fluctuations of the total particle number.
We first perform a numerical study where we optimize the Fisher information of the obtained state by exploring systematically  the parameter space for experimentally accessible configurations.
{\rien In contrast to} {\rien reference} \cite{Simon}, where the Husimi function of the macroscopic superposition was 
{\rien considered} ({\rien this distribution does not exhibit} fringes), we look at the interference fringes of the Wigner function to quantify the survival of quantum correlations in the presence of decoherence (see endnote \cite{endnote51}).
We also calculate the Fisher information of the state after averaging over many stochastic realizations in order to quantify its usefulness for metrology. The numerical study is followed by an analytical part that gives a limpid interpretation of the results, {\rien see section \ref{sec:interpret}.}

Finally in section \ref{sec:Maotcf} we give up the two-mode approximation for a truly multimode description {\rien of the bosonic system} and we estimate what are the constraints on the temperature {\rien of the 
Bose-condensed gas  used in} the preparation of the initial phase state, in order to obtain {\rien the desired}
mesoscopic superposition {\rien with a good fidelity} and a {\rien significant} revival in the phase contrast.
{\rien We conclude in section \ref{sec:conclusion}.}

\section{Cat-state fidelity versus contrast revival
	\label{sec:CLRA}}
In this section we show that there is a simple {\rien relation} between the fidelity of the state obtained at $t_{\rm cat}$ and the amplitude of the contrast revival peak at $t_{\rm rev}=2t_{\rm cat}$. 
{\rien For simplicity, we consider in this section two spatially separated components, with the same scattering length and loss rates, as one would have by using two symmetric Zeeman sub-levels as internal states, or by using two spatially separated BECs in the same internal state.}

We neglect fluctuations of the total particle number assuming that an initial state with a fixed number of particles can be prepared, for example by melting a Mott insulator phase in an optical lattice, {\rien or by non-destructive detection of the atoms with an optical cavity.}

{\rien Since the bosonic field populates two orthogonal modes with corresponding annihilation operators
$\hat{a}$ and $\hat{b}$, one can attribute an effective spin $1/2$ to the bosons and introduce the 
usual single-spin Bloch representation} and the usual {\rien dimensionless} collective spin operators \cite{Rev_Pezze}:
\begin{equation}
\hat{S}_x=\frac{\hat{a}^\dagger \hat{b}+\hat{b}^\dagger \hat{a}}{2} \;; \:
\hat{S}_y=\frac{\hat{a}^\dagger \hat{b}-\hat{b}^\dagger \hat{a}}{2i} \;; \:
\hat{S}_z=\frac{\hat{a}^\dagger \hat{a}- \hat{b}^\dagger \hat{b}}{2} \; .
\label{eq:spintotal}
\end{equation}
We consider an initial phase state with $N$ particles, on the equator of the Bloch sphere, 
with a relative phase $\varphi=0$ between the two modes:
	\begin{equation}
	\ket{\psi (0)}=  \frac{1}{\sqrt{N!}}\bb{ \frac{\hat{a}^{\dagger} + \hat{b}^{\dagger}}{\sqrt{2}}  }^N \ket{0}  \equiv \ket{\frac{\pi}{2};0}_N \,.
	\label{eq:phase0N}
	\end{equation}
Here the phase state with $N$ atoms is defined as
	\begin{equation}
	\ket{\theta ;\, \varphi}_N  \equiv \frac{1}{\sqrt{N!}}\left[ \left(\cos \frac{\theta}{2}\right) \: e^{i \frac{\varphi}{2}} \;\hat{a}^{\dagger} +  \left(\sin \frac{\theta}{2}\right) \: e^{- i\frac{\varphi}{2}}\; \hat{b}^{\dagger}  \right]^N \! \ket{0}  \,.
	\label{eq:phaseN}
	\end{equation}
The relative phase $\varphi \in [-\pi,\pi]$ has a meaning modulo $2\pi$ and {\rien the polar angle}
$\theta \in [0,\pi]$.	
The initial state {\rien (\ref{eq:phase0N})} evolves under the influence of a nonlinear spin Hamiltonian
{\rien resulting from the elastic $s$-wave interactions inside each mode \cite{DalibardCastin,Sinatra98}},
	\begin{equation}
	H={\rien \hbar}\chi \hat{S}_z^2 = \frac{{\rien\hbar}\chi}{2} \left( \hat{N}_a^2 + \hat{N}_b^2 - \frac{\hat{N}^2}{2}\right) \,,
	\label{eq:Sym_HSz2}
	\end{equation}
and in the presence of particle losses (one-, two- and three-body) within each {\rien spatial} component. 
The whole evolution is governed by the master equation for the density operator \cite{Sinatra98,LiYunSqueez}:
	\begin{equation}
	\frac{d}{dt}\hat{\rho}={\rien \frac{1}{i\hbar}}\left[\hat{H},\hat{\rho}\right]
	+ \mathcal{L}_1 {\rien [}\hat{\rho}{\rien]}+\mathcal{L}_2 {\rien[}\hat{\rho}{\rien]}
+\mathcal{L}_3{\rien[}\hat{\rho}{\rien]}\,  {\rien ,}
	\label{eq:master-eqn}
	\end{equation}
where {\rien the Liouvilian operators are} $\mathcal{L}_m = \mathcal{L}_m^{(a)} + \mathcal{L}_m^{(b)}$ with
	\begin{equation}
\mathcal{L}_m^{(a)} {\rien[}\hat{\rho}{\rien]}=\frac{1}{2}\gamma^{(m)}\bb{\left[\hat{a}^m,\hat{\rho}\bb{\hat{a}^{\dagger}}^m\right]
	+\left[\hat{a}^m\hat{\rho},\bb{\hat{a}^{\dagger}}^m\right] }
		\end{equation} 
and similarly for the mode $b$. {\rien Note that there are no collisions between modes $a$ and $b$ because they are spatially separated.} {\rien The rates $\gamma^{(m)}$ are related to the loss rate constants
$K_m$ and to the (in practice Gross-Pitaevskii) normalised condensate wavefunction $\phi(\rr)$ in one of the modes
by $m\gamma^{(m)}={\rien K_m} \int d^3r |\phi(\rr)|^{2m}$ \cite{LiYunSqueez}. The loss rate constants
are such that, in the spatially homogeneous zero-temperature Bose gas with $N$ particles
and mean density $\rho$, the $m$-body losses lead to a decay $\frac{d}{dt} N=-K_m \rho^{m-1} N$.}
		
In the absence of losses, at the time $t_{\rm cat}= \frac{\pi}{2\chi}$, the system  is in a Schr\"odinger cat state
{\rien given by
\begin{multline}
|\psi(t_{\rm cat}) \rangle = e^{-i \frac{\pi}{2} \hat{S}_z^2} \ket{\frac{\pi}{2};0}_N  \\ =
e^{i\frac{\pi}{8}(N^2-2)}e^{i\frac{\pi}{2}N \hat{S}_z} \bb{\frac{ \ket{\frac{\pi}{2};0}_N  + i e^{i\frac{\pi}{2}N}\ket{\frac{\pi}{2};\pi}_N }{\sqrt{2}} } \,.
\label{eq:chat_tourne}
\end{multline}
This results from the identity $\exp(-i\pi n^2/2)=\exp(i\pi/4)[\exp(i\pi n)-i]/\sqrt{2}$, for $n$ integer, and from the expansion of the initial state over Fock states. 
\be
|\psi(0) \rangle = \frac{1}{2^{N/2}} \sum_{N_a=0}^N \left(\frac{N!}{N_a! N_b!}\right)^{1/2} |N_a,N_b\rangle 
\label{eq:baleze}
\ee
with $N_b=N-N_a$.
Equation (\ref{eq:chat_tourne}) agrees with equation (19) in reference \cite{Ferrini2011} up to a global phase factor but it disagrees with reference \cite{Moelmer}. By using the relation 
$\exp(i \alpha \hat{S}_z)\ket{\frac{\pi}{2};\varphi}_N=\ket{\frac{\pi}{2};\varphi+\alpha}_N$, it can be rewritten as 
\begin{eqnarray}
\ket{\psi (t_{\rm cat})} & \stackrel{N \, \mathrm {even}}{=}& e^{-i\pi/4}\bb{\frac{\ket{\frac{\pi}{2};0}_N + i \ket{\frac{\pi}{2};\pi}_N}{\sqrt{2}}} \: \label{eq:cat_even} \\
\ket{\psi (t_{\rm cat})} &\stackrel{N \, \mathrm {odd}}{=}& e^{-i\pi/8}\bb{\frac{\ket{\frac{\pi}{2};\frac{\pi}{2}}_N -  \ket{\frac{\pi}{2};\frac{3\pi}{2}}_N}{\sqrt{2}}} .
\label{eq:cat_odd}
\end{eqnarray}
}
In presence of losses, we introduce the fidelity $\mathcal{F}(t)$ of the state {\rien $\hat{\rho}$} at time $t$, that is its overlap with the ``target'' state {\rien that} one would obtain in the lossless case:
\begin{equation}
\mathcal{F} (t) \equiv {\rm Tr}\{\hat{\rho}(t) |\psi^0 (t)\rangle\langle \psi^0 (t)| \}
\label{eq:Fidelity}
\end{equation}
The normalized first order correlation function between the two modes gives the contrast of the interference pattern if the two modes are made to interfere:
\begin{equation}
g^{(1)}(t)=\frac{2}{N} \langle \hat{S}_x \rangle(t)
\label{eq:g1}
\end{equation}
Its maximum value is one, realized at $t=0$ when the system is in a phase state.
In the lossless case $g^{(1)}(t)=\pm 1$ is recovered at multiples of the revival time $t_{\rm rev}=2t_{\rm cat}$.
We show here that for weak losses, one has to a very good approximation 
\begin{equation}
{\cal F}(t_{\rm cat}) = |g^{(1)}(t_{\rm rev})|^{1/2}
\label{eq:result_Fidelity}
\end{equation}

\subsection{Proof in the constant loss rate approximation}
The Monte Carlo wave function method \cite{MCWF,Belavkin,Barchielli} provides us with a stochastic formulation of the 
master equation (\ref{eq:master-eqn}). In this point of view the density matrix is seen as a statistical mixture of pure states $|\tilde{\psi}(t)\rangle$, each of which evolves under the influence of a non-hermitian effective Hamiltonian $H_{\rm eff}$ and {\rien of} random quantum jumps.
In terms of the jump operators $\hat{J}_\epsilon^m$ that annihilate $m$ particles in component $\epsilon=a,b$:
\begin{equation}
\hat{J}_a^m=\sqrt{\gamma^{(m)}}\hat{a}^m \,, \quad \quad \hat{J}_b^m=\sqrt{\gamma^{(m)}}\hat{b}^m
\end{equation} 
the effective Hamiltonian takes the form
\begin{equation}
H_{\rm eff}=H-\frac{i\hbar}{2} \sum_{\epsilon=a,b} \sum_{m=1}^{3} (\hat{J}_\epsilon^\dagger)^m \hat{J}_\epsilon^m \,.
\end{equation}
For a Monte Carlo wave function $| \tilde{\psi} (t)\rangle$ normalized to {\rien unity}, quantum jumps occur with a (total) rate 
$\sum_{\epsilon=a,b} \sum_{m=1}^{3} \langle \tilde{\psi}(t) | (\hat{J}_\epsilon^\dagger)^m \hat{J}_\epsilon^m|\tilde{\psi}(t)\rangle$.

The so-called ``constant loss rate approximation'', introduced in reference \cite{Sinatra98}, consists of the replacement
$(\hat{J}_\epsilon^\dagger)^m \hat{J}_\epsilon^m \to \gamma^{(m)} \bar{N}_\epsilon^m$ in the effective Hamiltonian, where 
$\bar{N}_\epsilon=N/2$ is the mean initial number of particles in each component.
Under this approximation, {\rien which} can be used when the mean fraction of lost particles is small, 
 the probability that $n$ quantum jumps have occurred at time $t$ is given by a Poisson law with parameter $\bar{n}=\lambda t$ where
\begin{equation}
\lambda=2\sum_{m=1}^{3} \gamma^{(m)} \bb{\frac{N}{2}}^m \,.
\label{eq:lambda}
\end{equation}
In this approximation the effective Hamiltonian indeed reduces to $H_{\rm eff}=H-\frac{i\hbar}{2} \lambda$ so that the probability that no jump occurs during a time delay $\tau$ is $||e^{-i H_{\rm eff}\tau/\hbar}|\tilde{\psi}\rangle||^2=e^{-\lambda \tau}$.

In expression (\ref{eq:Fidelity}) of the fidelity, only the realizations where no particles were lost at the cat-state time contribute.
In this subspace the density matrix evolves only under the influence of the effective Hamiltonian and remains in a pure state 
$P_N \hat{\rho}(t) P_N=e^{-\lambda t}|\psi_0(t)\rangle \langle \psi_0(t)|$, {\rien where $P_N$ projects onto the subspace with $N$ atoms.} The fidelity at the cat-state time is then 
\be
\mathcal{F}(t) = e^{-\lambda t_{\rm cat}} \,.
\label{eq:simple_fidelity}
\ee

On the other hand, we have shown in reference \cite{Sinatra98} that the peak in the contrast at the revival time in the presence of losses is 
$g^{(1)}(t_{\rm rev})={\rien (-1)^N} e^{-\lambda t_{\rm rev}}={\rien (-1)^N} e^{-2\lambda t_{\rm cat}}$, 
which concludes the proof.
A more detailed analysis, beyond the constant loss rate approximation, is performed analytically {\rien and numerically} in Appendix \ref{app:Losses} for one- and three-body losses (see endnote \cite{endnote52}).
{\rien We show there that, in the interesting regime in which the number of lost atoms at the revival time is smaller than one, $m \lambda t_{\rm rev}<1$  (the fidelity and the revival would be killed by the losses otherwise), the relative correction 
to (\ref{eq:result_Fidelity})
\begin{equation}
\frac{|{\rien |g^{(1)}(t_{\rm rev})|} -  {\rien {\cal F}(t_{\rm rev}/2)}^2|}{ {\rien {\cal F}(t_{\rm rev}/2)}^2} \approx 
\frac{2}{(m\pi)^2} \bb{m \lambda t_{\rm rev} \times  \frac{m \lambda t_{\rm rev}}{N} }
\label{eq:corr}
\end{equation}
can be interpreted, up to a factor $2/(m\pi)^2$, as the product between the number of lost atoms and the fraction of lost atoms at the revival time. It is hence $\ll1$.}

\subsection{Numerical example}	
	\label{sec:example}	
\begin{figure}[htp]
	\begin{center}
			\includegraphics[scale=1]{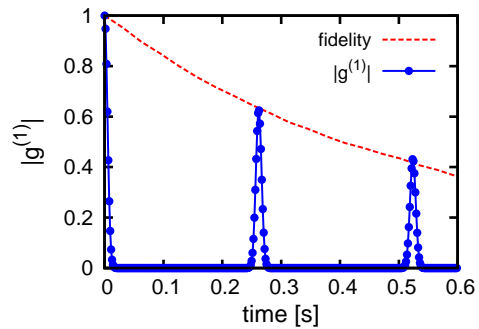}
		\end{center} 
		\caption{Fidelity and {\rien absolute value of} 
contrast versus time for  {\rien a split Bose-Einstein condensate of $N = 300$ {\rien $|F=1,m_F=-1\rangle$} ${}^{87}{\rm Rb}$ atoms in two identical and spatially separated harmonic potentials {\rien in the presence of} three-body losses. } Scattering length {\rien $a =100.4 a_0$,}
		trapping frequency $\omega/2\pi =500$ Hz, {\rien with a three-body loss constant rate}
		{\rien $K_3 = 6\times 10^{-42} $m$^6$/s \cite{Egorov}.}
		This gives {\rien $\chi =  12$ s$^{-1}$ and $\gamma^{(3)} = 2.6 \times 10^{-7} $} s$^{-1}$.
		\label{fig:experiment}}
	\end{figure}

	We show in Fig.~\ref{fig:experiment} an example where we solve numerically the master equation  in the presence of three-body losses  
(see endnote \cite{endnote53}) and compare the evolutions of the $g^{(1)}$ function and of the fidelity, confirming that the relation (\ref{eq:result_Fidelity}) approximately holds also beyond the constant loss rate approximation. 
The height of the first revival {\rien peak} is {\rien $0.63$ and the fidelity of the cat state {\rien is} $0.79$}. 

The conclusion of this section is twofold. First, losses should be limited to less than one particle on average at the cat-state time to preserve a high fidelity. Second, we have shown that there is a simple quantitative relation  (\ref{eq:result_Fidelity}) between the amplitude of the revival peak in the contrast and the cat-state fidelity.
The physical reason is that each loss event introduces a random shift of the relative phase between the modes (see section \ref{sec:interpret}), {\rien corresponding to} a rotation of the state around the $z$ axis, by an angle of order $\chi t$ where $t$ is the time at which the loss occurred. As $\chi t$ is of the order of $\pi$ at the cat-state time or the revival time, one particle lost on average is sufficient to kill both the cat state and the phase revival.
	
\section{{\protect \rien Realistic analysis for rubidium or sodium atoms on a} hyperfine transition}
\label{sec:HTWROSA}

{\rien This section gives} {\rien a description of the two-mode dynamics {\rien as close as possible to the experimental state of the art,}} including losses and particle number fluctuations.  The two condensed modes 
correspond to two different atomic internal sub-levels, already used and coupled in cold atom experiments
by a hyperfine transition. As $N$ fluctuates, we take a different perspective on the cat-state formation:
the goal is no longer to prepare with highest fidelity the pure cat state (\ref{eq:cat_even}) or (\ref{eq:cat_odd}), it is rather 
to produce a mixed cat state with maximal usefulness for precision measurements,
that is maximal Fisher information. The ``catiness'' of the mixed state is revealed by fringes in
the Wigner distribution function.

\subsection{Experimental constraints}
\begin{figure}[htb]
	\includegraphics[scale=0.45]{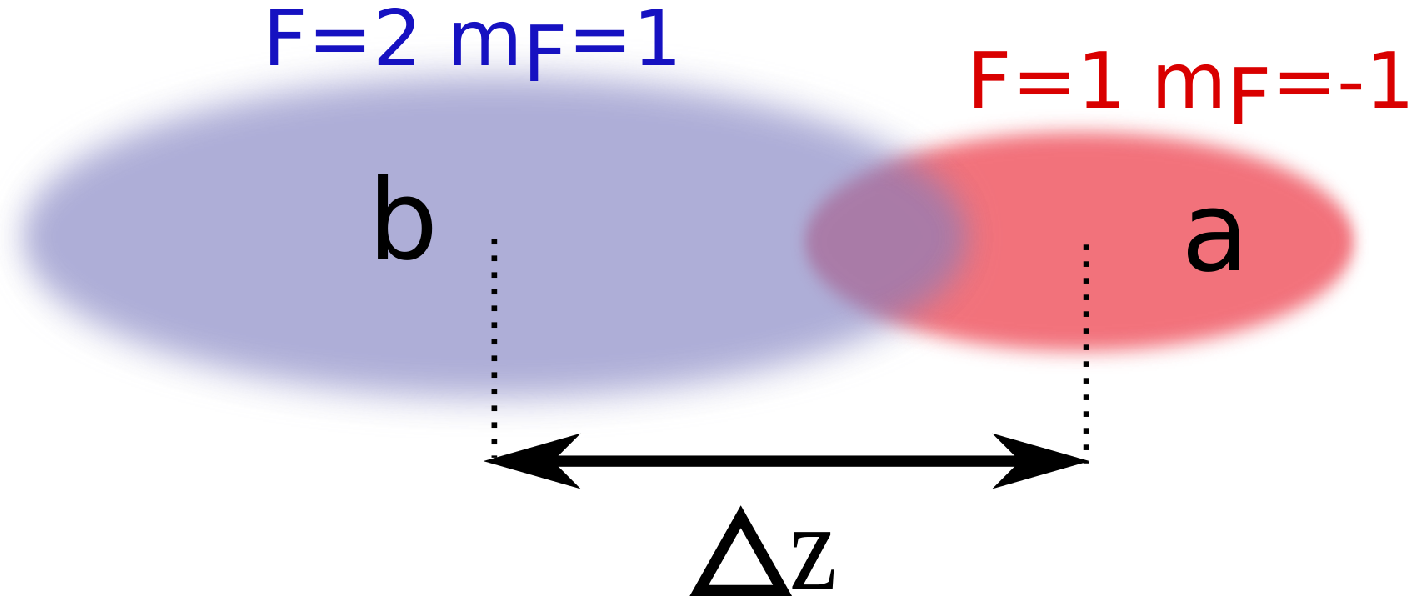} \\ \vspace{0.5cm}
	\includegraphics[scale=0.45]{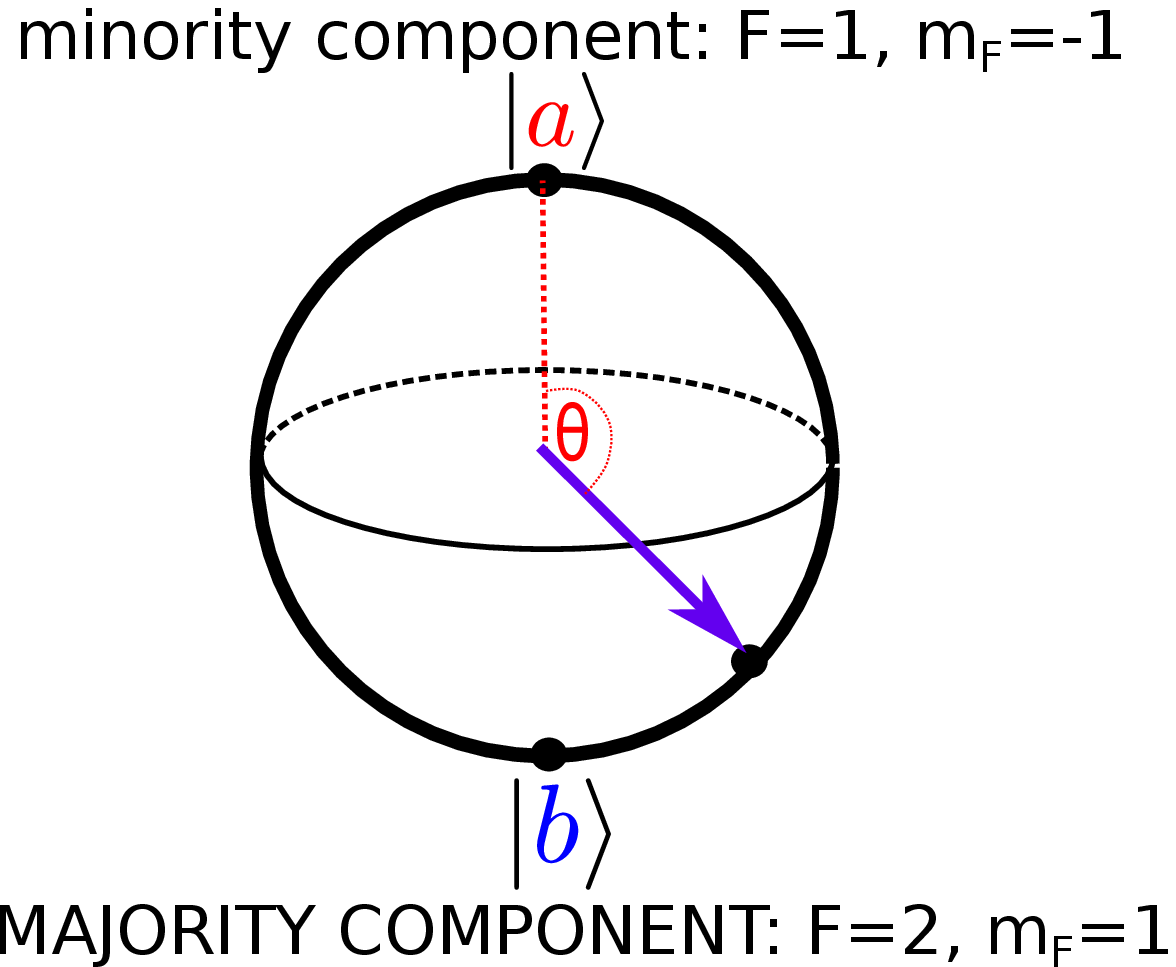}	
	\caption{Top: {\rien Trapping} configuration for rubidium 87 atoms: two cigar-shaped harmonic traps displaced by $\Delta z$ along the ``long'' {\rien trap} axis.
		Bottom: Representation of the initial state on the Bloch sphere: $\ket{b} = \ket{F=2, m_F=1}$ is the majority component and $\ket{a} = \ket{F=1, m_F=-1}$ is the minority component spin state. 
		The initial state is close to the south pole ($\theta$ {\rien close to $\pi$}).
		\label{fig:scheme}}
\end{figure}
We {\rien now} concentrate on the two internal states of rubidium 87 $|F=1,m_F=-1\rangle$ and $|F=2,m_F=1\rangle$ that have been used to generate spin squeezing in state dependent potentials on a chip \cite{SqMunich,MWpotentials}. The experimental constraints that we consider are  
(i) large two-body losses in $|F=2,m_F=1\rangle$ due to spin changing collisions,
(ii) limited {\rien background} lifetime (we take {\rien $1/K_1=5$} s) in both states due to imperfect vacuum,
(iii) fluctuations of the total number of atoms $N$.
{\rien 
Concerning this last effect, we remark that, even in the absence of losses,  the orientation of 
the cat state depends on $N$ modulo 4. This is apparent from equation (\ref{eq:chat_tourne}) where a $N$-dependent rotation around the $z$-axis acts on a state (the state between parentheses) with $N$-independent coefficients in the Fock basis.} 
If $N$ fluctuates {\rien with a standard deviation $\gg 1$ as in regular experiments}, the interference fringes at the cat-state time are then completely washed out when averaging over $N$. In  these conditions one might think that it would be difficult, if not impossible, to create a cat state under the experimental constraint mentioned above. We will show that this is not the case. However, in order to counteract decoherence we will have to consider a more general situation than the one described in Section \ref{sec:CLRA}. We will (i) de-symmetrize the initial mixture by performing a large pulse instead of a $\pi/2$-pulse, (ii) de-symmetrize the two trapping potentials, and (iii) allow for an overlap between the two {\rien spatial} modes. This is schematized in Fig.~\ref{fig:scheme}.

After rubidium we consider the two internal states of sodium 23 $|F=1,m_F=0\rangle$ and $|F=2,{\rien m_F=-2}\rangle$ {\rien in more general, cigar-shaped or pancake-shaped state-dependent potentials}.
In this case spin changing collisions {\rien between atoms} in $F=2$ are suppressed and {\rien $a$-$b$ losses are negligible   
 \cite{Na_scatt}. The losses can then} be significantly 
{\rien lowered} provided a very good vacuum is achieved, which allows {\rien us} to push {\rien up} further the atom number in the quantum superposition. 

\subsection{\rien Numerical calculations}
\label{sub:num}

{\rien We first performed a numerical study to determine 
the optimal  experimental conditions} within the given constraints. 

The system state is supposed to be initially in a statistical mixture of phase states 
	\begin{equation}
		\hat{\rho}(0) = \sum_{N=0}^{\infty} p(N) \; |\theta ;\, \varphi\rangle_N {}_N\langle \theta ;\, \varphi| 
		\label{eq:rho0}
	\end{equation}
where the phase state $|\theta; \varphi \rangle_N $ with $N$ atoms is given in (\ref{eq:phaseN}), and
$p(N)$ is the distribution of the total number of atoms, assumed to be Poissonian of average $\bar{N}$. 

{\rien The master equation obeyed by $\hat{\rho}(t)$} is still of the form of Eq.\eqref{eq:master-eqn}, but with
non-symmetric $m$-body loss rates
$\gamma_\epsilon^{(m)}$ for $m=1,2,3$ and $\epsilon=a,b$:
\begin{eqnarray}
\gamma_\epsilon^{(m)} &=& \frac{K_\epsilon^{(m)}}{m} \int d^3r \; |\phi_\epsilon(r)|^{2m} \,,
\label{eq:nonsym_gamma}\\
\gamma_{ab} &=& \frac{K_{ab}}{2} \int d^3r \; |\phi_a(r)|^{2} |\phi_b(r)|^{2} \,,
\label{eq:nonsym_gamma_ab}
\end{eqnarray}
where $K_\epsilon^{(m)}$ and $K_{ab}$ are loss rate constants, {\rien and $\gamma_\epsilon^{(m)}$ and $\gamma_{ab}$
are calculated using the stationary normalized condensate wave functions $\phi_\epsilon(r)$ for} $N_a=\bar{N}_a$ and $N_b=\bar{N}_b$. As now the modes can spatially overlap, we also include two-body
processes, with rate $\gamma_{ab}$, where one atom in $a$ and one atom in $b$ are lost at the same time
{\rien \cite{Egorov}} (see endnote \cite{endnote54}).

The {\rien unitary} part of the evolution {\rien in the master equation} is calculated {\rien with} the zero-temperature mean-field model {\rien Hamiltonian}
	\begin{equation}
	H_{\rm \rien GP} = \sum_{N_a, N_b =0}^{\infty} E_{\text{\rien GP}}(N_a, N_b) \ket{N_a, N_b}\bra{N_a, N_b} 
	\label{eq:hamGPE}
	\end{equation}
	 where $E_{\text{\rien GP}}$ is the Gross-Pitaevskii energy 
	\begin{eqnarray}
	E_{\text{\rien GP}}(N_a, N_b) &=& \sum_{\epsilon=a, \,b}  N_{\epsilon} \left[ \int \phi_{\epsilon}^* h_{\epsilon} \phi_{\epsilon}
	+ \frac{g_{\epsilon \epsilon}}{2}N_\epsilon \int |\phi_\epsilon|^4 \right]
	 \nonumber \\
	&+&  g_{ab}N_aN_b\int |\phi_a|^2 |\phi_b|^2 {\rien \;.}
	\end{eqnarray}
The single particle Hamiltonians {\rien $h_a$ and $h_b$} include the kinetic energy and the trapping potential.
The {\rien stationary condensate wave functions $\phi_{\epsilon}$ and the} Gross-Pitaevskii energy $E_{\rm \rien GP}$ have been computed numerically for different pairs $(N_a,\,N_b)$ (in practice a few thousands) to construct the Hamiltonian \eqref{eq:hamGPE}.

{\rien In order to find the optimal conditions, we were scanning the experimental parameters space, each time performing the evolution starting from the initial condition (\ref{eq:rho0}),} optimizing entanglement {\rien witnesses} that are sensitive to {\rien the presence of a Schr\"odinger} cat. {\rien To avoid extreme parameters that would make the experimental realization more difficult,
we have restricted the search to trapping frequencies ratios smaller than 20.}
Details of our procedure are given in Appendix \ref{app:search}, {\rien and two examples of results are shown in the next subsection}.

\subsection{Fisher information and Wigner function of the cat state}

\begin{figure}[htb]
	\begin{center}
		\includegraphics[width=0.4\textwidth]{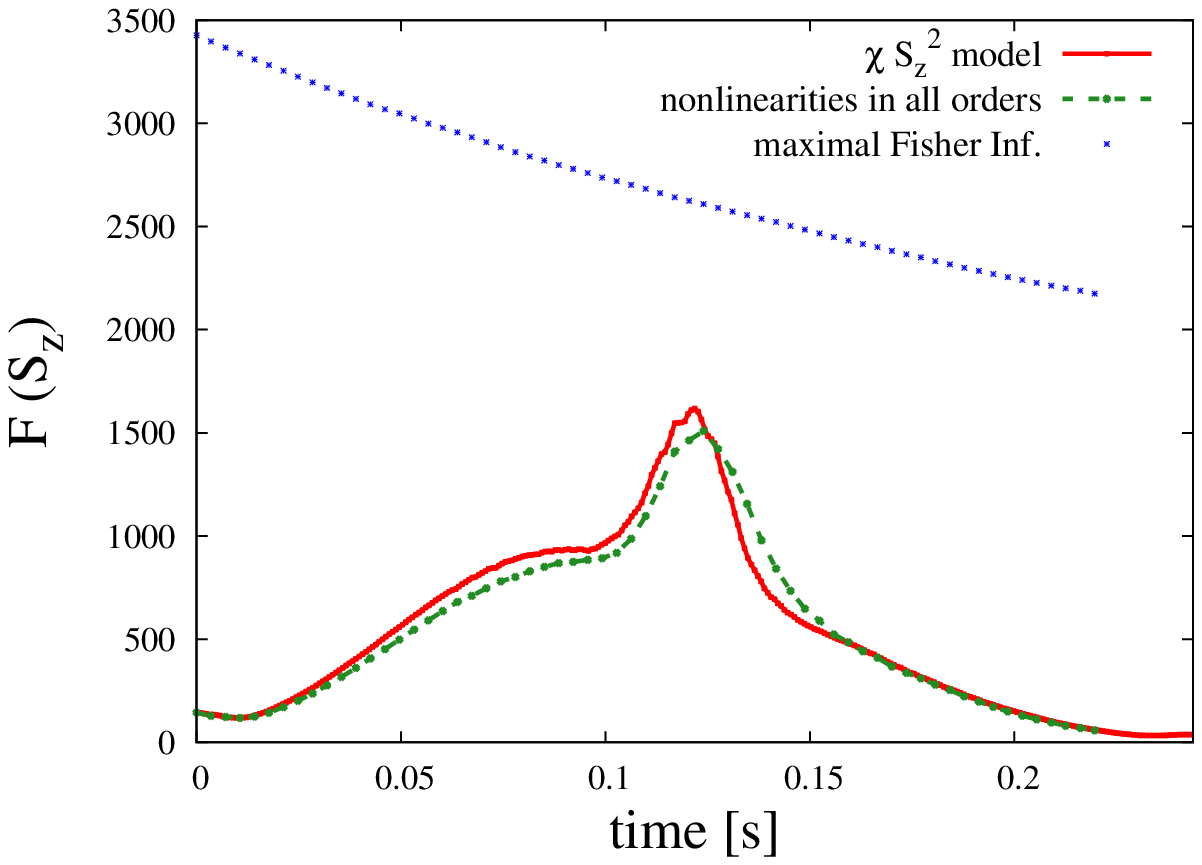}
		\hspace{0.0cm}
		\includegraphics[width=0.4\textwidth]{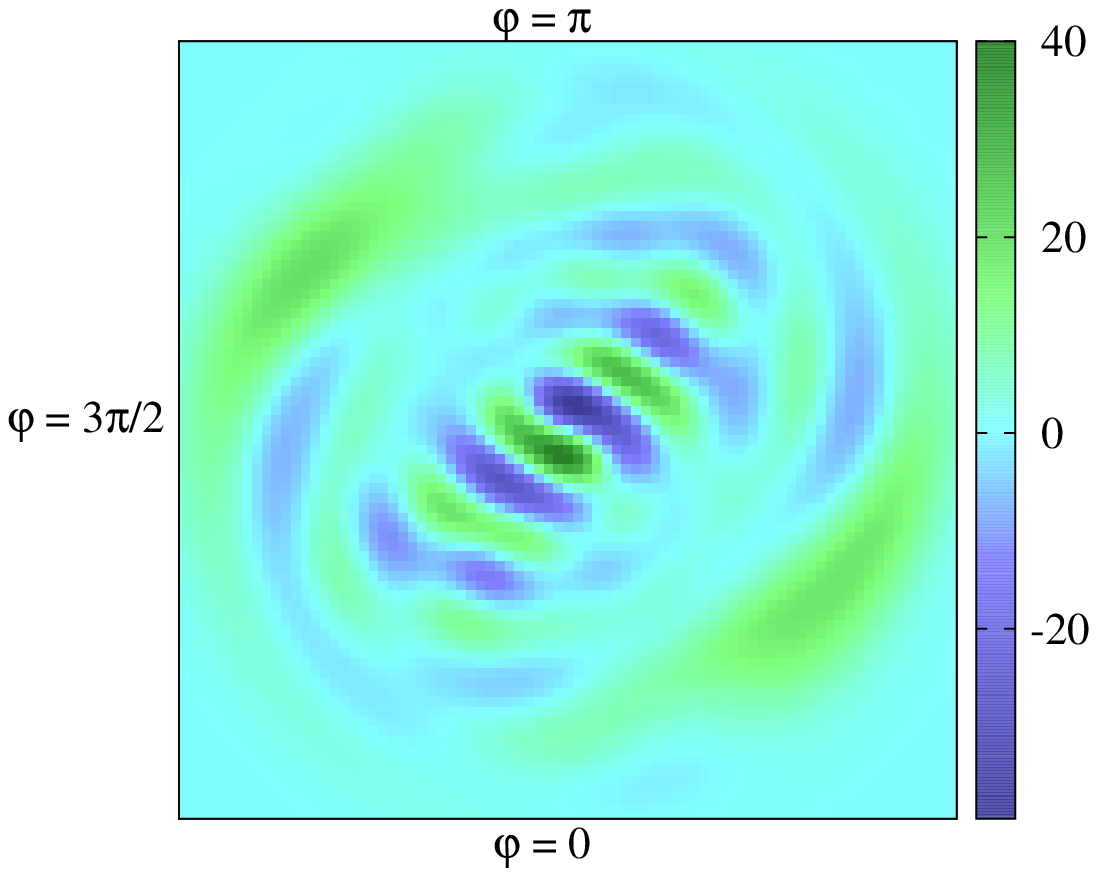}
	\end{center} 
	\caption{
{\rien Optimal cat state that we predict for realistic} experimental conditions {\rien with} the two rubidium {\rien 87} states  $|a\rangle=|F=1,m_F=-1\rangle$ and $|b\rangle=|F=2,m_F=1\rangle$.
Top: Fisher information (\ref{eq:fisher}) as a function of time, calculated with the {\rien Gross-Pitaevskii} Hamiltonian (\ref{eq:hamGPE}) {\rien (green {\rien dash-dotted} line)}, and with the general two-mode model of section \ref{subsec:gtmm} {\rien (red solid line)}; {\rien for comparison, the blue {\rien dotted} curve {\rien equation (\ref{eq:Fishertheta})} gives the maximal Fisher information that one {\rien could obtain}
 for the {\rien time-dependent mean atom numbers}.} {\rien Bottom:}  Wigner function at $t_{\rm cat}=112$ ms calculated with the {\rien Gross-Pitaevskii} Hamiltonian (\ref{eq:hamGPE}).
{\rien The Wigner function, in the south hemisphere of the Bloch sphere, is projected onto the $x-y$ plane.}   
Parameters: {\rien The total atom number of average $\bar{N}=150$ has Poissonian fluctuations}, $\bar{N}_a=5.71$, $\bar{N}_b=144.29$, $\omega_\perp=2\pi \times 1000$ Hz, 
$\omega_{za}=2\pi \times 850$ Hz, $\omega_{zb}=2\pi \times 50$ Hz, $\Delta z=1.620 \, a_{\rien \perp}$ 
(distance between the trap centers). Scattering lengths {\rien $a_{aa}=100.4 a_0$, $a_{bb}=95.44 a_0$, $a_{ab}=98.13 a_0$ \cite{Egorov}}. 
One-body, two-body, and three-body loss rate constants $K_a^{(1)}=K_b^{(1)}=0.2\, {\rm s}^{-1}$, $K_b^{(2)}=8.1\times10^{-20} \, {\rm m}^3/{\rm s}$, $K_{ab}=1.51\times10^{-20} \, {\rm m}^3/{\rm s}$, $K_a^{(3)}=6\times10^{-42} \, {\rm m}^6/{\rm s}$ {\rien \cite{Egorov}}. {\rien In the} general two-mode model of section \ref{subsec:gtmm}, these parameters {\rien lead} to  $\chi=12.895 \, {\rm s}^{-1}$, $\tilde{\chi}=12.888 \, {\rm s}^{-1}$, $\gamma_{b}^{(2)}=6.436\times10^{-3}\, {\rm s}^{-1}$, $\gamma_{ab}=1.032 \times10^{-3}\, {\rm s}^{-1}$, $\gamma_{a}^{(3)}=5.15\times10^{-6}\, {\rm s}^{-1}$.} 
	\label{fig:wigS}
\end{figure}
\begin{figure}[htb]
	\begin{center}
		\includegraphics[width=0.4\textwidth]{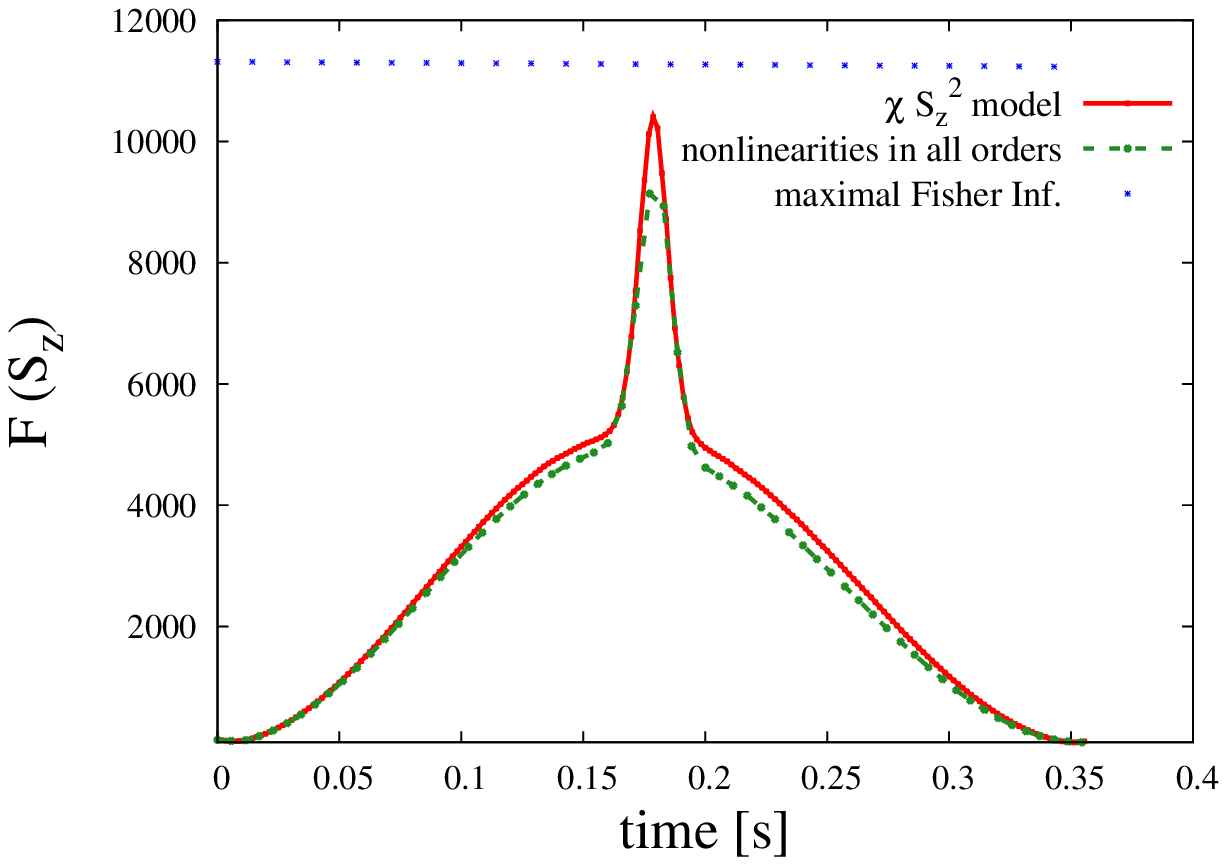}
		\hspace{0.0cm}
		\includegraphics[width=0.4\textwidth]{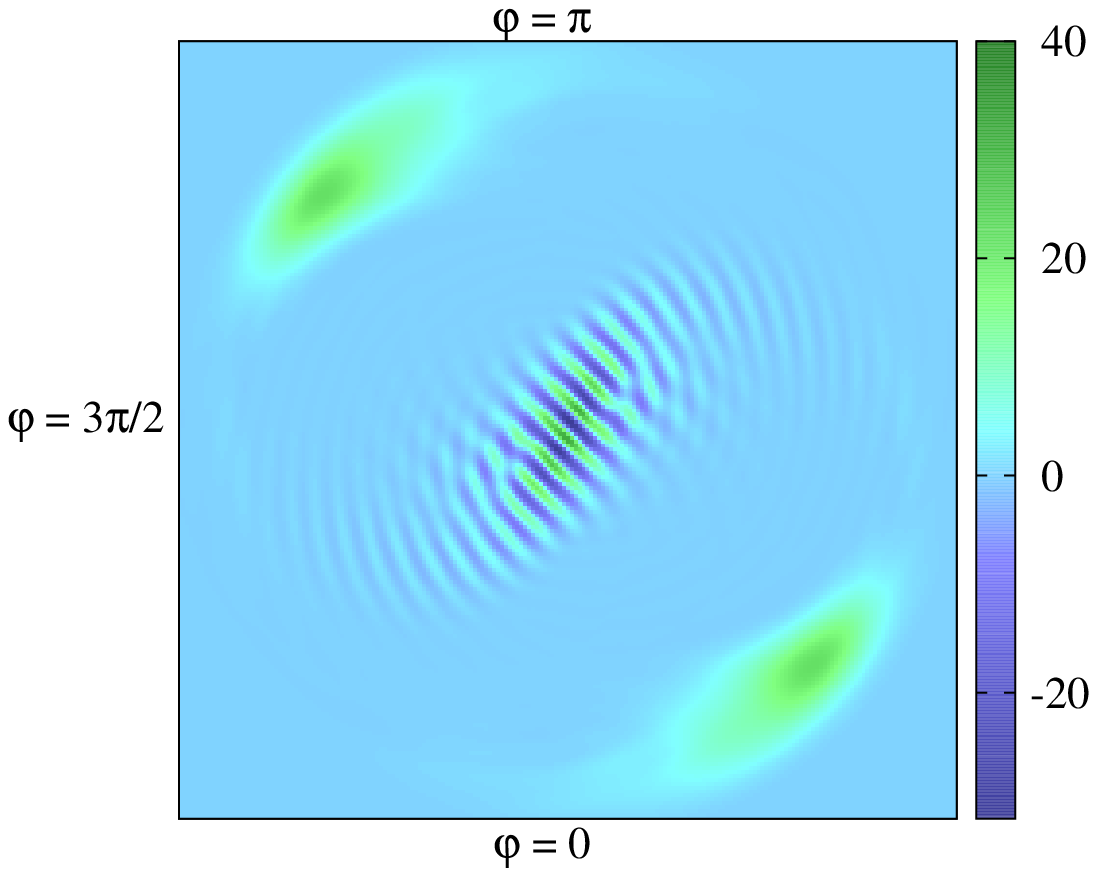}
	\end{center} 
	\caption{
{\rien Optimal cat state that we predict for realistic} experimental conditions {\rien with} the two sodium 
{\rien 23} states  $|a\rangle=|F=1,m_F=0\rangle$ and $|b\rangle=|F=2,m_F=-2\rangle$.
Top: Fisher information (\ref{eq:fisher}) as a function of time, calculated with the {\rien Gross-Pitaevskii} Hamiltonian (\ref{eq:hamGPE}) {\rien (green {\rien dash-dotted} line)}, and with the general two-mode model of section \ref{subsec:gtmm} {
\rien (red solid line)}; {\rien for comparison, the blue {\rien dotted} curve {\rien equation (\ref{eq:Fishertheta})} gives the maximal Fisher information that one {\rien could obtain} for the {\rien time-dependent mean atom numbers}.} Bottom:  Wigner function at $t_{\rm cat}=178$ ms calculated with the {\rien Gross-Pitaevskii} Hamiltonian (\ref{eq:hamGPE}). 
{\rien The Wigner function, in the south hemisphere of the Bloch sphere, is projected onto the $x-y$ plane.}  
Parameters: {\rien The total atom number of average $\bar{N}=150$ has Poissonian fluctuations}, $\bar{N}_a=22$, $N_b=128$, $\omega_{\perp a}=2\pi \times 1415$ Hz, $\omega_{\perp b}=2\pi \times 115$ Hz, $\omega_{za}=2\pi \times 612$ Hz, $\omega_{zb}=2\pi \times 772$ Hz, $\Delta z=0.62\times10^{-6}$m (distance between the trap centers). Scattering lengths {\rien $a_{aa}=52.91 a_0$, $a_{bb}=64.25 a_0$, $a_{ab}=64.25 a_0$} \cite{Na_scatt}. One-body loss rate constants $K_a^{(1)}=K_b^{(1)}=0.01\, {\rm s}^{-1}$. {\rien In the} general two-mode model of section \ref{subsec:gtmm}, these parameters {\rien lead} to $\chi=8.763 \, {\rm s}^{-1}$, $\tilde{\chi}=8.729 \,{\rm s}^{-1}$. {\rien We expect no relevant $a$-$b$ or $b$-$b$ two-body losses here \cite{Na_scatt}}, and we have checked that for the considered parameters the contribution of three-body losses is negligible.}
	\label{fig:wigS_Na}
\end{figure}

{\rien For optimized conditions issued by our search algorithm (see Appendix \ref{app:search}),}
in Fig.~\ref{fig:wigS} and  Fig.~\ref{fig:wigS_Na} we show the {\rien resulting} time evolution of the Fisher information, and the Wigner distribution at the cat-state time, obtained {\rien respectively} for rubidium {\rien 87} {\rien and} sodium {\rien 23,} {\rien for} the hyperfine transitions mentioned above. 

\begin{figure}[htb]
	\begin{center}
	\includegraphics[width=0.25\textwidth]{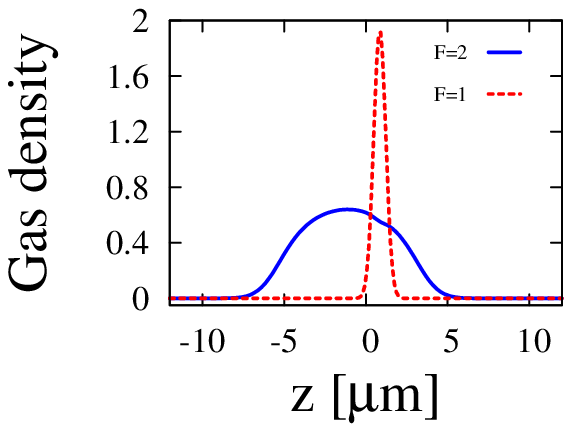}\includegraphics[width=0.25\textwidth]{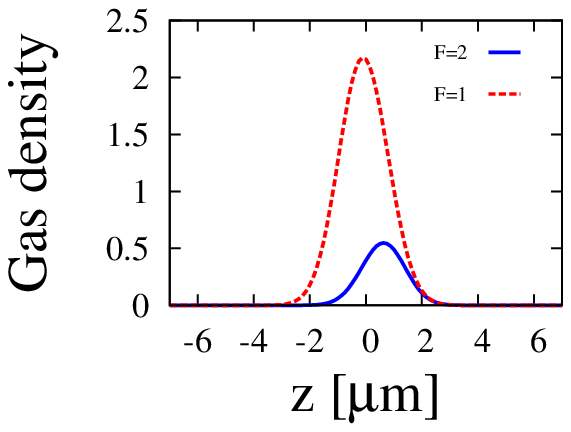}
	\end{center} 
	\caption{
Density cuts along the $z$-axis of the {\rien majority} (blue {\rien solid line}) and the {\rien minority} {\rien (red {\rien dashed line})} component for parameters of Fig.~\ref{fig:wigS} for rubidium {\rien 87} (left), and of Fig.~\ref{fig:wigS_Na} for sodium {\rien 23} (right).
	} 
	\label{fig:dens}
\end{figure}

The corresponding {\rien cuts through the atomic density distribution} 
 along the $z$-axis {\rien of the trap} for the two states are shown in Fig.~\ref{fig:dens}.
The Fisher information $F(\Sz)$ {\rien that we plot} quantifies the sensitivity of the state to a small rotation {\rien around} axis $\bm{n}$ lying in the equator of the Bloch sphere, when a {\rien measurement} of {\rien the observable $\hat{S}_z$} is performed:
	\begin{eqnarray}
	F(\Sz) &=& {\rm Max}_{\:\bm{n}}  F(\Sz, \bm{n}) \,, \label{eq:fisher} \\
	F(\Sz, \bm{n}) &=& \lim_{\alpha\to 0} \sum_{k=0}^{\infty}\frac{1}{p(k | \alpha,\bm{n})} 
	\bb{ \frac{d\,p(k | \alpha,\bm{n})}{d\alpha} }^2  \,,
	\end{eqnarray}
where $p(k|\alpha , \bm{n})$ is the probability of finding $k$ atoms in the {\rien minority} component $a$
in the rotated state $\hat{\rho}_{\alpha} = e^{-i \alpha \hat{\bf S}\cdot {\bm{n}}}  \hat{\rho} e^{ i \alpha \hat{\bf S}\cdot {\bm{n}}}$.
{\rien We have chosen $\hat{S}_z$ as the observable with respect to which {\rien we} define the Fisher information, because one can show that in the ideal lossless case, even for a large pulse as in Fig.~\ref{fig:scheme}, $F(\Sz)$ reaches the quantum Fisher information obtained by maximizing $F(\hat{O}, \bm{m})$ both with respect to the measured observable $\hat{O}$ and to the rotation axis  ${\bm{m}}$ of the state.} {\rien In an experiment one has to consider in addition the finite resolution of the atom number counting in the modes $a$ and $b$. If the detection system does not quite reach single atom resolution, one can still detect the cat state and determine the Fisher information if one chooses a spin observable in the $x-y$ plane, oriented along the direction of the fringes in Figs.~3 and 4 respectively.} 

The {\rien optimized} results in Fig.~\ref{fig:wigS} and  Fig.~\ref{fig:wigS_Na} include Poissonian fluctuations of the total particle number, finite lifetime and particle losses for both states, {\rien that is one-body losses and, for ${}^{87}$Rb, three-body and two-body losses including inter-component $a$-$b$ losses}. 

{\rien In the lossless case, the maximal Fisher information achievable when starting with an initial phase state (\ref{eq:phaseN}) with a total atom number $N$ and a pulse angle $\theta$ is
\begin{equation}
F(\Sz) = N^2 \sin^2 \theta + N\cos^2 \theta \,.
\label{eq:Fishertheta}
\end{equation}
In Fig.~\ref{fig:wigS} and  Fig.~\ref{fig:wigS_Na} we show as a function of time this maximal Fisher information
with $N=N_a(t)+N_b(t)$ and $\theta=\arccos \frac{N_a(t)-N_b(t)}{N_a(t)+N_b(t)}$ corresponding to the time-dependent mean atom numbers in our system in the presence of losses.}
{\rien It is} remarkable that, although $30$ particles are lost on average in the majority component in Fig.~\ref{fig:wigS} {\rien  for ${}^{87}$Rb atoms} (see Appendix \ref{app:search}), high contrast fringes are obtained in the Wigner function at the cat-state time, and {\rien the} corresponding Fisher information is reduced by a factor less than half with respect to its maximal possible value with the {\rien same} number of atoms.  {\rien From equation (\ref{eq:Fishertheta}) it is apparent that a non-symmetric pulse $\theta\neq \frac{\pi}{2}$ reduces the maximal Fisher information in the lossless case. The situation is however different in the presence of losses, where  the best pulse angle, as well as the best trap parameters can only be derived from an optimization procedure that is specific to the selected transition and the atomic species plus the experimental constraints (see Appendix \ref{app:search})}.

\section{Physical interpretation {\protect \rien of the realistic-analysis optimum}}
\label{sec:interpret}

{\rien We provide here a simple physical interpretation of the mixed cat state with maximal Fisher information 
numerically determined in section \ref{sec:HTWROSA} for experimentally realistic conditions including losses and
particle number fluctuations.}

\subsection{{\rien Analytical model in the general case}}
\label{subsec:gtmm}
{\rien As in the numerical simulations of section \ref{sub:num},
in the general non-symmetric case,  we use a master equation of the form \eqref{eq:master-eqn}, with non-symmetric $m$-body loss rates (\ref{eq:nonsym_gamma}) and (\ref{eq:nonsym_gamma_ab}), and with an initial condition 
(\ref{eq:rho0}) representing a statistical mixture of phase states with Poissonian fluctuations of the total atom number with average $\bar{N}$.
The difference here, in order to perform an analytical study and develop some intuition, is that the unitary evolution of the density matrix is not calculated with the fully nonlinear Hamiltonian (\ref{eq:hamGPE}) but with a non-symmetric {\rien $S_z^2$} Hamiltonian \cite{LiYun2009}}
\begin{equation}
	\hat{H}=  {\rien \hbar}\tilde{\chi} \hat{N}\, \Sz +{\rien \hbar} \chi \Sz^2 \,,
\label{eq:H}
\end{equation}
where we omitted terms that give a constant phase drift or a global phase shift. 
The microscopic expressions of $\tilde{\chi}$ and $\chi$  are \cite{LiYun2009}
	\begin{eqnarray}
	\chi&=&\frac{1}{2\hbar}\left( \partial_{N_a}\mu_a + \partial_{N_b}\mu_b - \partial_{N_a}\mu_b -\partial_{N_b}\mu_a   \right)_{\bar{N}_a,\bar{N}_b} \,, \\
	\tilde{\chi}&=&\frac{1}{2\hbar}\left( \partial_{N_a}\mu_a  -\partial_{N_b}\mu_b   \right)_{\bar{N}_a,\bar{N}_b} \,,
	\end{eqnarray}
where $\mu_a$ and $\mu_b$ are the chemical potentials of the $a$ and $b$ condensate respectively.
We calculate $\chi$ and $\tilde{\chi}$ by solving the stationary two-component Gross-Pitaevskii equation in the considered {\rien trap} geometry for different atom numbers around the averages $\bar{N}_a = \bar{N} \cos^2{\theta/2}$ and $\bar{N}_b = \bar{N} \sin^2{\theta/2}$,
while the loss parameters (\ref{eq:nonsym_gamma}) and (\ref{eq:nonsym_gamma_ab}) are calculated {\rien as mentioned earlier} for $N_a=\bar{N}_a$ and $N_b=\bar{N}_b$.

{\rien As can be noted from Fig.~\ref{fig:wigS} and Fig.~\ref{fig:wigS_Na} (see the captions),}
for all the {\rien optimal} 
configurations found by our algorithm, one has $|(\chi-\tilde{\chi})/(\chi+\tilde{\chi})|\ll 1$. We explain here the reason.

 \subsection{Compensation of random phase shifts}
 \label{sec:compensation}
By evolving an initial phase state $|\theta;0\rangle_N$ with the Hamiltonian \eqref{eq:H} in the absence of losses,
a cat state appears at the time $\tcat=\frac{\pi}{2\chi}$. 
We define the ``unrotated cat'' with $N$ particles as 
{\rien 
\begin{equation}
\ket{ {\rm unrot \; cat}}_N \equiv \frac{1}{\sqrt{2}}\bb{\ketph{\theta}{0}_N  + i \, e^{i\frac{\pi}{2}N} \ketph{\theta}{\pi }_N} \;.
\label{eq:cat0}
\end{equation}
}
By using the result for the evolution with {\rien a pure} $\chi \Sz^2$ Hamiltonian \cite{Ferrini2011} and by {\rien including the effect of the additional} $N$-dependent drift term {\rien $\hbar\tilde{\chi} \hat{N}\hat{S}_z$} in the Hamiltonian, we obtain at the cat-state time:
\be
\ket{\psi (\tcat)}_{N} = e^{i \hat{N}\frac{\pi}{2} \Sz(1-\frac{\tilde{\chi}}{\chi})}\ket{ {\rm {\rien unrot} \; cat}}_N.
\label{eq:psi_tcat}
\ee
This shows that for $\tilde{\chi}=\chi$ the dependence on $N$ of the cat orientation, {\rien impossible to avoid when $\tilde{\chi}=0$ as in equation (\ref{eq:chat_tourne}),} is now eliminated.

Let us now consider the effect of one-body losses (with {\rien a} rate $\gamma$) in one of the two components. 
Starting from a phase state, the trajectory with one atom lost at time $t_1$ in mode $a$ or $b$, can be {\rien expressed} in terms of a Hamiltonian evolution starting from a state with initially $N-1$ atoms plus a random $t_1$-dependent shift of the relative phase:
	\begin{equation}
		|\tilde{\psi}_{1\: a,b} (t) \rangle= {\cal N}_{a,b}
		{\color{black}	\underbrace{e^{-i \, \bb{\tilde{\chi} \pm \chi} \Sz \,t_1}}_{\text{$t_1$-dependent drift}}  }  			e^{-i H t/{\rien \hbar}} \ket{\psi (t=0)}_{N-1}
	\label{eq:1-jump-ab}
	\end{equation}
where ${\cal N}_{a,b}$ includes a global phase and a normalization factor, and the plus or minus sign in the $t_1$-dependent shift refers to a loss in component $a$ or $b${\rien ,} respectively.	
This shows that the random shift due to losses that comes from the quantum jump and from the $N$-dependent drift velocity in \eqref{eq:H}, can be set to zero in one of the two components by adjusting $\tilde{\chi}$ to $\mp \chi$, the two effects compensating each other.
In particular, for $\tilde{\chi}=\chi$ both the random shift due to losses in $b$ and the deterministic $N$-dependent rotation of the cat state that is present even without losses in (\ref{eq:psi_tcat}) are suppressed \cite{Krzysiek1}.
This conclusion, based here on the analysis of a conditional state with a single lost atoms (\ref{eq:1-jump-ab}), holds also in the case of two- and three-body losses \cite{Krzysiek2}.

We now have to distribute the roles of {\rien $a$-mode and $b$-mode} to the two hyperfine states of ${}^{87}$Rb $|F=1,m_F=-1\rangle$ and
$|F=2,m_F=1\rangle$. The key point is that {\rien the dominant loss process is two-body losses in} $|F=2,m_F=1\rangle$.
These are the ones that should be compensated. {\rien In addition} there will be unavoidable one-body losses in the majority component. All the significant losses should be concentrated in a single component where they can be compensated. This {\rien explains} the (at first sight) counterintuitive choice of taking $|b\rangle=|F=2,m_F=1\rangle$ as the majority component, {\rien done} in Fig.~\ref{fig:scheme}.

\subsection{Coherent state description}
A particularly simple interpretation of our results is obtained in the coherent state description that we will adopt in this subsection. To this aim we note that a Poissonian mixture of phase states for two modes is identical to a statistical mixture of {\rien Glauber} coherent states with random total phase and a fixed relative phase
\begin{eqnarray}
	\hat{\rho} &=& \sum_N \frac{\bar{N}^Ne^{-\bar{N}}}{N!} |\theta; \varphi \rangle_{N} {}_{N} \langle \theta; \varphi | \nonumber \\
	 &=& \int_0^{2\pi} \frac{d\Theta}{2\pi} |\alpha,\beta\rangle \langle \alpha,\beta|
	\label{eq:rho}
\end{eqnarray}
where $|\alpha,\beta\rangle$ is a two-mode coherent state $\alpha=\sqrt{\bar{N}_a}e^{i\varphi_a}$ and $\beta=\sqrt{\bar{N}_b}e^{i\varphi_b}$, with $\varphi=\varphi_a-\varphi_b$ the relative phase between the coherent states, $\Theta=\frac{\varphi_a+\varphi_b}{2}$ the total phase and $\bar{N}=\bar{N}_a+\bar{N}_b$ the {\rien mean} total atom number.
To show the equality (\ref{eq:rho}) one {\rien expands} the phase states and the coherent states over Fock states $|N_a,N_b\rangle$. The integral over $\Theta$ suppresses coherences between Fock states with different total numbers of particles.

{\rien The next step is to remark} that the Hamiltonian (\ref{eq:H})
can be {\rien elegantly} written {\rien as the sum} of two independent Hamiltonians {\rien plus a term that depends on $\hat{N}$ only,}
\begin{eqnarray}
	\hat{H}&=&\frac{{\rien\hbar}\chi_a}{2} \hat{N}_a^2 + \frac{{\rien\hbar}\chi_b}{2}  \hat{N}_b^2 - {\rien\hbar}\chi \frac{\hat{N}^2}{4}	\,, \label{eq:H2} \\
	\chi_a&=&\chi+\tilde{\chi} \quad ; \quad  \chi_b=\chi-\tilde{\chi}\,,  \label{eq:corr_chi}
\end{eqnarray}
despite the fact that the two modes overlap and interact with each other.
The term that depends on $\hat{N}$ only is irrelevant because there are no coherences between states of different $N$.
{\rien For each state} $|\alpha,\beta\rangle$ {\rien appearing in the statistical mixture (\ref{eq:rho}), the evolution of $a$ and $b$ modes under the influence of the Hamiltonian \eqref{eq:H2} and of losses {\rien other than $a$-$b$ losses}, is decoupled}.

{\rien \subsubsection{Evolution of the coherent states in the presence of losses}
In the {\rien remainder} of this section we consider the evolution of the two-mode coherent state $|\alpha,\beta\rangle$ under the influence of the Hamiltonian \eqref{eq:H2} and {\rien one-body} losses. Although strictly speaking these states are not physical and the integral in (\ref{eq:rho}) randomizing the total phase should be taken into account, the analysis gives some insight into the compensation condition, and it allows to introduce a fidelity that is not trivially zero in a case in which the total number of particles is not fixed.}

\medskip
{\rien \it {Perfect compensation case}} -
{\rien Let} us consider the effect of one-body losses {\rien first} in {\rien the case} $\tilde{\chi}=\chi$ {\rien that is} $\chi_b=0$. In section \ref{sec:compensation} we refer to this condition as ``compensation'' because in the Monte Carlo wave function approach, the random phase shifts coming from the losses and the $N$-dependent drift of the relative phase compensate.
After the transformation \eqref{eq:H2} we can call it as well ``no effective interactions in $b$''. In this case, even in the presence of losses, the state of mode $b$ remains a pure state: it is an exponentially decreasing coherent state
\begin{equation}
	|\psi_b(t)\rangle=|\tilde{\beta}\rangle \quad \mbox{where} \quad \tilde{\beta}=\beta e^{-\frac{\gamma_b}{2}t} \,.
\end{equation}
This can be seen in the {\rien Monte} Carlo wave function method, where, after renormalisation, we obtain $|\tilde{\psi}_b(t)\rangle=|\tilde{\beta}\rangle$ for any quantum trajectory evolving under the influence of the non-hermitian Hamiltonian $H_{\rm eff} = -\frac{i\hbar}{2} 
\gamma_b b^\dagger b$ and $k$ jumps with jump operator $C=\sqrt{\gamma_b}b$.
Since the mode $b$ {\rien is effectively non-interacting ($\chi_b=0$)}, it constitutes a perfect phase reference even in the presence of losses. Only its amplitude decreases in time. A similar conclusion was already reached in references \cite{Krzysiek1,Krzysiek2}.

In the absence of losses in $a$, $\gamma_a=0$, with $\chi_a\neq0$, the mode $a$ evolves as described {\rien in reference \cite{Yurke}}, going through a Schr\"odinger cat at time $t_{\rm cat}=\frac{\pi}{\chi_a}=\frac{\pi}{2\chi}$ and a revival at time $t_{\rm rev}=\frac{2\pi}{\chi_a}=\frac{\pi}{\chi}$.
In particular, for $|\psi_a(0)\rangle=|\alpha\rangle$, we have
\begin{equation}
	|\psi_a^0(t_{\rm cat}=\frac{\pi}{\chi_a})\rangle = \frac{1}{\sqrt{2}} \left[ e^{-i\frac{\pi}{4}}|\alpha\rangle + e^{i\frac{\pi}{4}}|-\alpha\rangle \right]
\end{equation}
and
\begin{equation}
	|\psi_a^0(t_{\rm rev}=\frac{2\pi}{\chi_a})\rangle =|-\alpha\rangle \,.
\end{equation}
The {\rien exponent} on $\psi_a^0$ recalls that this is the ideal, {\rien lossless} case in mode $a$.

What happens in the presence of {\rien one-body} losses {\rien of rate $\gamma_a$} in mode $a$ ?
Something close to a cat state can only be obtained if these losses are very weak (less than one atom lost on average at the cat-state time). This means $|\alpha|^2\gamma_a t_{\rm cat}<1$ and hence $\gamma_a t_{\rm cat} \ll 1$.
Within the Monte Carlo wave function approach, we introduce the non-normalized state vector $|\tilde{\psi}_a(t)\rangle$, corresponding to a trajectory for mode $a$ where no atoms were lost in that mode at time $t$. Noting 
that the {\rien effective non-hermitian} Hamiltonian {\rien can be written in a form equivalent to 
(\ref{eq:H2})}, as the sum of commuting parts, and introducing $\hat{H}_a\equiv {\rien\hbar}\chi_a \hat{N}_a^{\rien 2}/2$, we have
\be
|\tilde{\psi}_a(t)\rangle= {\rien e^{-\gamma_a \hat{N}_a t/2} e^{-\frac{i}{\hbar}\hat{H}_a t}| {\rien \alpha} \rangle=}
\tilde{A}e^{-\frac{i}{\hbar}\hat{H}_a t}| \tilde{\alpha}\rangle
\ee
with
\be
\tilde{\alpha}=\alpha e^{-\frac{\gamma_a}{2}t} \quad \mbox{and}\quad 
\tilde{A}=e^{-\frac{|\alpha|^2}{2}} e^{\frac{|\tilde{\alpha}|^2}{2}} \,.
\ee
In the coherent state {\rien description}, and before taking the integral over $\Theta$, we define the fidelity of the state {\rien resulting} from the evolution with losses as
\begin{equation}
	{\cal F}\equiv |\langle \tilde{\psi}_a(t) |\psi_a^0(t) \rangle |^2 \,.
\end{equation}
From the previous equations, at the cat-state time (neglecting for $|\alpha|\gg 1$ the vanishing overlap $\langle -\alpha|\alpha \rangle$) one then has
\begin{equation}
	{\cal F}(t_{\rm cat})=|\tilde{A}\langle \alpha|\tilde{\alpha} \rangle|^2=|e^{-|\alpha|^2(1-e^{-\frac{\gamma_a}{2}t_{\rm cat}})}|^2 \simeq e^{-|\alpha|^2 \gamma_a t_{\rm cat}}
	\label{eq:Fc_kk}
\end{equation}
and similarly at the revival time
\begin{equation}
	{\cal F}(t_{\rm rev}) \simeq e^{-|\alpha|^2 \gamma_a t_{\rm rev}} {\rien \simeq} ({\cal F}(t_{\rm cat}))^2  \,.
	\label{eq:Fr_kk}
\end{equation}
Let us now look at the relative amplitude of the revival peak of the normalized $g^{(1)}$ function. For $|\alpha|^2\gg1$ we obtain:
\begin{equation}
	g^{(1)}(t_{\rm rev}) = \frac{\langle \hat{a} \rangle(t_{\rm rev})}{\langle \hat{a} \rangle(0) }
\simeq  {\rien -} e^{-|\alpha|^2 \gamma_a t_{\rm rev} } \,,
	\label{eq:kk}
\end{equation}
showing that the amplitude of the revival peak directly gives informations on the cat-state fidelity
\begin{equation}
	|g^{(1)}(t_{\rm rev})| =({\cal F}(t_{\rm cat}))^2 \,.
\end{equation}
This is again the relation (\ref{eq:result_Fidelity}), this time for coherent states and in the more general asymmetric case.

{\rien {\it \small Note -The fact that one can restrict to the zero-loss subspace to define the cat-state fidelity is less clear
when the target state is a coherent superposition of Glauber coherent states: the action of the jump
operator $\hat{a}$ describing the loss of a particle does not render the coherent state $|\alpha\rangle$
orthogonal to itself (contrarily to the case of states with well defined particle numbers). The zero-loss
subspace restriction performed in {\rien equations (\ref{eq:Fc_kk})-(\ref{eq:kk})} is however exact for the defined quantities $\mathcal{F}(t_{\rm cat})$,
$\mathcal{F}(t_{\rm rev})$ and $\langle \hat{a}\rangle(t_{\rm rev})$ in the limit $\gamma_a t_{\rm cat}\to 0$
at fixed $|\alpha|^2\gamma_a t_{\rm cat}$.  This can be checked from the exact expressions, obtained
using equations (6) and (7) of reference \cite{LiYunSqueez} to calculate the density operator $\hat{\rho}_a$ of mode $a$ in
presence of one-body losses:
\begin{eqnarray}
\mathcal{F}(t) &=& \langle \psi_a^0(t)|\hat{\rho}_a(t)|\psi_a^0(t)\rangle \nonumber \\
&=&\sum_{k\in\mathbb{N}}\frac{\gamma_a^k}{k!} \int_{[0,t]^k} dt_1\ldots dt_k e^{-\gamma_a \sum_{j=1}^k t_j} \nonumber \\
 \times&&\!\!\!\!\!\!\!\!\!\!\! \exp\left\{-2|\alpha|^2 \left[1-e^{-\gamma_a t/2}\cos(\chi_a\sum_{j=1}^k t_j)\right]\right\}
\end{eqnarray}
\begin{eqnarray}
 \langle\hat{a}\rangle(t_{\rm rev}) &=& \mbox{Tr}[\hat{a}\hat{\rho}_a(t_{\rm rev})] \nonumber  \\
 &=& - e^{-\gamma_a t_{\rm rev}/2} e^{-|\alpha|^2(1-e^{-\gamma_a t_{\rm rev}})} \nonumber  \\
&\times & \exp\left[\frac{\gamma_a}{\gamma_a+i\chi_a}\left(1-e^{-\gamma_a t_{\rm rev}}\right)\right]       
\end{eqnarray}
For the fidelity,
one is helped by the fact that, in this large $|\alpha|^2$ limit, the randomness of the particle-loss
time, combined with the evolution with the quartic Hamiltonian $\hat{H}_a$, effectively
results at times of order $1/\chi_a$ into a {\sl large} random phase shift of the coherent state amplitude $\alpha$.}}

What is actually measured in an experiment is $|\langle ab^\dagger \rangle|$, where the expectation value is taken in (\ref{eq:rho}) and the integral over $\Theta$ must be performed. This experimental contrast then reads
\begin{equation}
	{\cal R}_{\rm exp}\equiv |\langle \hat{a} \hat{b}^\dagger \rangle|(t_{\rm rev}) \simeq 
	|\alpha| e^{-|\alpha|^2 \gamma_a t_{\rm rev} } |\beta|e^{-\frac{\gamma_b}{2}t_{\rm rev}} \,.
\end{equation}
If the fraction of atoms lost in $b$ at $t=t_{\rm rev}$ is small, then $e^{-\frac{\gamma_b}{2}t_{\rm rev}}\simeq1$ and one essentially recovers (\ref{eq:kk}).

\medskip
{\rien \it {Imperfect compensation of the lossy mode}} - 
If $\chi_b \ll \chi_a$ but $\chi_b\neq0$, there are some residual effective interactions in the mode $b$. As a consequence our phase reference starts to undergo a phase collapse. This modifies the contrast as follows:
\begin{equation}
	{\cal R}_{\rm exp} \simeq 
	|\alpha| e^{-|\alpha|^2 \gamma_a t_{\rm rev} } |\beta|e^{-\frac{\chi_b^2 |\beta|^2 t_{\rm rev}^2}{2}}
\end{equation}
plus small corrections due to losses in $b$. 
The compensation constraint becomes stringent for large atom numbers as one must have 
$ \chi_b / \chi_a  \ll 2 / \sqrt{ \pi N_b  } $.
If no compensation is done at all, that is $\chi_b\simeq\chi_a$, there will be no revival at all in 
${\cal R}_{\rm exp}$. Indeed as the mode $b$ is lossy with $|\beta|^2 \gamma_b t_{\rm rev}>1$  it has a phase collapse with no revival. 

\section{Multimode analysis of the cat-state formation: nonzero temperature effects}
\label{sec:Maotcf}

In this paper, up to now, we have analysed the quantum dynamics of the bosonic field in a two-mode model. Reality is however multimodal,
and there is always a nonzero thermal component in the initial state of the system, which can endanger the cat-state production even in the absence
of losses. {\rien In this section we discuss} nonzero temperature effects, both on the cat-state fidelity and on the contrast
revival, in the Bogoliubov approximation.

\subsection{Proposed experimental procedure}

In the multimode case, one must revisit the definition of the initial state (\ref{eq:phase0N}) and explain how to prepare it. In order to avoid any excitation induced by the $\pi/2$ pulse (see equation (28) in reference \cite{EPL}), we assume as in reference \cite{KurkjianGP} that the gas is initially non-interacting, $g_{aa}(0^-)=g_{ab}(0^-)=g_{bb}(0^-)=0$,
and prepared at thermal equilibrium at the lowest accessible temperature $T$ with all the $N$ bosons in internal state $|a\rangle$.
At time zero, to obtain a phase state, one applies an instantaneous $\pi/2$ pulse between the $|a\rangle$ and $|b\rangle$ states, which transforms the atomic field operators in the Heisenberg picture as follows:
\bea
\label{eq:apresa}
\hat{\psi}_a(\rr,0^+) &=& \frac{1}{\sqrt{2}} \left[\hat{\psi}_a(\rr,0^-)-\hat{\psi}_b(\rr,0^-)\right] \\
\label{eq:apresb}
\hat{\psi}_b(\rr,0^+) &=& \frac{1}{\sqrt{2}} \left[\hat{\psi}_a(\rr,0^-)+\hat{\psi}_b(\rr,0^-)\right]
\eea
To obtain the nonlinear spin dynamics required to get a cat state, one adiabatically increases the interaction strength $g_{aa}(t)=g_{bb}(t)=g(t)$ up to the final value $g_{\rm f}$ in a time $t_{\rm ramp}$, while keeping $g_{ab}=0$, and one lets the system evolve until the much longer cat-state production time $t_{\rm cat}$ or contrast revival time $t_{\rm rev}$.

{\rien {\it \small Note -- Experimentally, to suppress interactions, one can start with a condensate at low enough atomic density, perform the $\pi/2$ pulse and spatially separate the components $a$ and $b$. The total number of Bogoliubov excitations created by the pulse in each component  in the homogeneous case, for $g_{aa}(0^-)=g_{bb}(0^-)=4\pi \hbar^2 a(0^-)/m$ and $g_{ab}=0$, is $N_\sigma^{\rm exc}(0^+) \simeq 0.395 N\,\sqrt{\rho a(0^-)^3}$ (from equation (38) and Appendix C of reference \cite{Casagrande}) and should be $\ll1$.
The condition $g_{ab}=0$ is ensured by spatial separation of the $a$ and $b$ components right after the pulse using state dependent potentials \cite{MWpotentials,CiracZoller_linanglelin}. Note that the interaction dynamics are much slower than the 
$\pi/2$ pulse and the subsequent spatial separation.
Once the components are split, the effective interaction strength is increased by adiabatically reducing the volume of the trapping potentials of the two components. Alternatively, an atomic species with a Feshbach resonance in state $|a\rangle$ could be used, which allows tuning the interaction strength $g_{aa}=0$ \cite{Inguscio}. The $\pi/2$ pulse could then be performed in real space (rather than on the spin degrees of freedom) with all atoms in $|a\rangle$ by adiabatically ramping up a barrier in the trapping potential to split the atomic cloud. Subsequently, the Feshbach resonance is used to tune the interactions in both wells of the resulting double well potential to a nonzero value. Finally, if one prefers to avoid barrier splitting and interaction suppression by decompression, a possibility is to use spin-1 bosonic particles, with $|a\rangle$ and $|b\rangle$ the internal states of maximal spin $\pm \hbar$ along
the quantization axis $Oz$ as for example $|F=1,m_F=\pm 1\rangle$. The spinor symmetry then imposes equal coupling constants $g_{aa}=g_{bb}$ in the two states. Unfortunately, the internal scattering lengths
of $|F=1,m_F=\pm 1\rangle$ are expected to have a magnetic Feshbach resonance at opposite values $\pm B_0$ of the magnetic field along the quantization axis $Oz$. 
Generically one thus cannot achieve
$g_{aa}=g_{bb}=0$ for a given value of such a magnetic field $B \mathbf{e}_z$. 
A first solution is to make $B$ rapidly oscillate in time between opposite values such that on average
$g_{aa}=g_{bb}=0$. A second solution is to rapidly and coherently transfer back the $b$ atoms into the internal state $a$ after the $\pi/2$ pulse and the spatial separation of the two spin components, {\sl e.g.} with a spatially resolved laser-induced Raman transition. -- }}

 For simplicity, and taking into account recent experiments on degenerate gases in
flat bottom potentials \cite{Hadzibabic_uniforme,Zwierlein_uniforme}, we assume in this section that each spin component is trapped in a cubic box of volume $V=L^3$ with periodic boundary conditions. One can then take advantage of the fact that the Bogoliubov mode functions are plane waves
with known amplitudes, which makes explicit calculations straightforward. As an immediate illustration, we give an adiabaticity
condition for the interaction switching in Appendix \ref{appen:adiab}, for the Hann ramp
\be
g(t)=\frac{g_{\rm f}}{2} \left(1-\cos \frac{\pi t}{t_{\rm ramp}}\right) \ \ \ \mbox{for}\ 0<t<t_{\rm ramp}\,.
\label{eq:Hann}
\ee

\subsection{Analysis at zero temperature} In the ideal limit of $T=0$, the system is initially prepared in its ground state,
with the $N$ bosons in internal state $|a\rangle$ with a vanishing wavevector $\kk=\mathbf{0}$.  Just after the $\pi/2$ pulse, 
due to (\ref{eq:apresa},\ref{eq:apresb}), the system is in the state
\be
|\psi(0^+)\rangle =\frac{1}{(N!)^{1/2}2^{N/2}} [\hat{c}_{a,\mathbf{0}}^\dagger(0^-)+\hat{c}_{b,\mathbf{0}}^\dagger(0^-)]^N |0\rangle
\ee
where the bosonic operator $\hat{c}_{\sigma,\kk}$ annihilates a particle in internal state $|\sigma\rangle$ with wavevector $\kk$
and $|0\rangle$ is the vacuum. The binomial expansion gives
\begin{multline}
|\psi(0^+)\rangle = \frac{1}{2^{N/2}} \sum_{N_a=0}^{N} \left(\frac{N!}{N_a!N_b!}\right)^{1/2}  \\ \times |N_a:a,\kk=\mathbf{0}; N_b: b,\kk=\OO\rangle \,.
\label{eq:2m_evolue}
\end{multline}
In this form, each Fock state is the ground state of the system at the considered fixed values of $N_a$ and $N_b=N-N_a$. Under adiabatic
switching of the interaction strength, it is transformed into the instantaneous ground state $|\psi_0(N_a,N_b;t)\rangle$
of the interacting system (taken with a real wavefunction), with instantaneous energy $E_0(N_a,N_b;t)$. The global state of the system
is then at time $t$:
\begin{multline}
|\psi_{\rm adiab}(t)\rangle =  \frac{1}{2^{N/2}} \sum_{N_a=0}^{N} \left(\frac{N!}{N_a!N_b!}\right)^{1/2} 
\\ \times  e^{-i \int_0^t d\tau E_0(N_a,N_b;\tau)/\hbar}  |\psi_0(N_a,N_b;t)\rangle \,.
\label{eq:fond_evolue}
\end{multline}
This defines the equivalent of the phase state and its evolution in the multimode theory. In the large $N$ limit, one recovers a $\hat{S}_z^2$ 
spin dynamics as in Eq.~(\ref{eq:Sym_HSz2}) by expanding $E_0(N_a,N_b;\tau)$ around $(\bar{N}_a,\bar{N}_b)=(N/2,N/2)$ up to second order
in $N_a-\bar{N}_a=-(N_b-\bar{N}_b)=(N_a-N_b)/2$. At $t>t_{\rm ramp}$,  this gives ({\rien see the note in the next paragraph})
\begin{multline}
|\psi_{\rm adiab}(t)\rangle \simeq  \frac{e^{-i\int_0^td\tau E_0(\bar{N}_a,\bar{N}_b;\tau)/\hbar}}{2^{N/2}}
e^{-i\chi\hat{S}_z^2 (t-t_0)} \\ \times \sum_{N_a=0}^{N} \left(\frac{N!}{N_a!N_b!}\right)^{1/2} |\psi_0(N_a,N_b;t)\rangle
\label{eq:psimulti}
\end{multline}
with the collective spin operator $\hat{S}_z=(\hat{N}_a-\hat{N}_b)/2$ and the spin nonlinearity coefficient
\be
\chi = \frac{1}{\hbar} \frac{\partial^2 E_0}{\partial{N_\sigma^2}} (\bar{N}_a,\bar{N}_b;g=g_{\rm f})
\ee
{\rien where $\sigma$ is any of the $a$, $b$.}
The phenomenology of cat-state formation and contrast revival of the two-mode model is straightforwardly recovered, up to a 
retardation time $t_0$ due to the adiabatic ramping of the interaction,
\be
\int_0^t d\tau \frac{1}{\hbar} \frac{\partial^2 E_0}{\partial{N_\sigma^2}} (\bar{N}_a,\bar{N}_b;\tau) \stackrel{t>t_{\rm ramp}}{=} \chi (t-t_0)
\ee
{\rien The pure state (\ref{eq:psimulti}), 
and the resulting cat state at the appropriate time, exhibits {\rien entanglement}
between the external orbital degrees of freedom and the internal spin degrees of freedom. 
This {\rien entanglement} can be eliminated by adiabatically ramping down the interaction strength to zero,
to transform back each $|\psi_0(N_a,N_b;t)\rangle$ into the Fock state 
$|N_a:a,\kk=\mathbf{0}; N_b: b,\kk=\OO\rangle$ with spin-state independent orbital modes.}

{\rien {\it \small Note -- If one expands $E_0(N_a,N_b;\tau)$ in equation (\ref{eq:fond_evolue}) up to fourth order in $N_a-\bar{N}_a=-(N_b-\bar{N}_b)$ in the spirit of figures \ref{fig:wigS} and \ref{fig:wigS_Na} (red curve vs green curve), one finds a state $|\psi_{\rm quart}(t)\rangle$ that differs from the state $|\psi_{\rm quad}(t)\rangle$ resulting from the second order expansion (as given by Eq.~(\ref{eq:psimulti})), because the Bogoliubov ground-state energy of the uniform gas is not purely quadratic in $N$ (contrarily to the Gross-Pitaevskii approximation). At the first time $t_{\rm cat}$ where $|\psi_{\rm quad}(t)\rangle$ is the target cat state, we find an overlap of the form $\langle \psi_{\rm quart}(t_{\rm chat})|\psi_{\rm quad}(t_{\rm chat})\rangle = {}_N\langle \pi/2;\varphi=0| \exp(i\alpha \hat{S}_z^4) |\pi/2;\varphi=0\rangle_N$
with $\alpha\simeq \frac{\pi}{24} \frac{\partial_{N_\sigma}^4E_0(\bar{N}_a,\bar{N}_b;g=g_{\rm f})}{\partial_{N_\sigma}^2E_0(\bar{N}_a,\bar{N}_b;g=g_{\rm f})}$. 
In the phase state, $\langle \hat{S}_z^4\rangle=N(3N-2)/16$.
The small parameter controlling the expansion is thus $\varepsilon={}_N\langle \pi/2;\varphi=0|\alpha \hat{S}_z^4|\pi/2;\varphi=0\rangle_N= -3 (2\pi)^{1/2} (\rho a_{\rm f}^3)^{1/2}/48$ in the thermodynamic limit, where $\rho=N/L^3$ is the total density. For the parameters of Fig.~\ref{fig:g1T} 
we find the  very small value $\epsilon\simeq -0.001$. This legitimates the quadratic expansion of $E_0$
for the uniform gas. In reality, cubic box potentials correspond to hard walls rather than to periodic boundary conditions. For the parameters of
Fig.~\ref{fig:g1T}, the healing length in a given spin component $\xi_\sigma=\hbar/(m\mu_\sigma)^{1/2}$ is significantly smaller than the box size $L$,
so to calculate the Gross-Piatevskii chemical potential $\mu_\sigma$,
we use the approximate condensate wavefunction $\phi_\sigma(\rr)=[\mu_\sigma/(N_\sigma g_{\rm f})]^{1/2}\prod_{\alpha=x,y,z} \tanh(r_\alpha/\xi)\tanh[(L-r_\alpha)/\xi]$,
knowing that the hyperbolic tangent form is exact for a single wall. The normalisation of $\phi_\sigma$ to unity leads to the equation of state
in the box $N_\sigma=\mu_\sigma(L-2\xi_\sigma)^3/g_{\rm f}$. We obtain $\xi_\sigma/L\simeq 0.17$ and $\epsilon\simeq 0.01$, which again validates
the quadratisation of $E_0$. -- }}

\subsection{Fidelity at nonzero temperature} 
\label{subsec:fideltot}

\begin{figure}[htb]
\begin{center}
\includegraphics[width=0.45\textwidth,clip=]{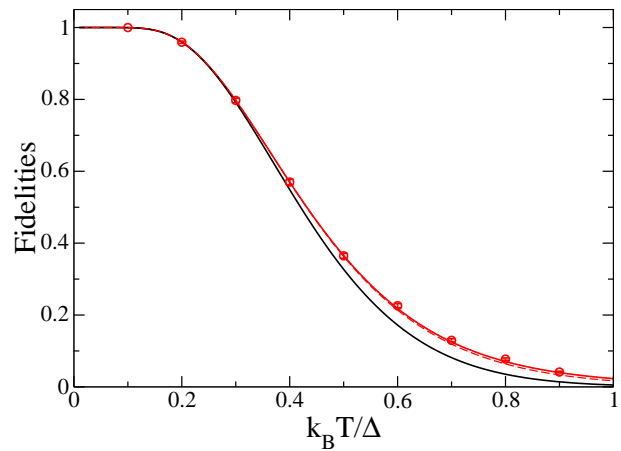}	
\end{center}
\caption{{\rien Fidelities} of the multimode cat-state preparation as a function of the temperature $T$ of the initial ideal gas
in the canonical ensemble, in a cubic box of size $L$ with periodic boundary conditions. 
{\rien Black (lower) solid line: fidelity $\mathcal{F}$ of the orbito-spinorial cat state
as given by equation (\ref{eq:fidelcan}). In red: peak fidelity $\mathcal{F}_{\rm spin}$ of the spin cat state as defined by equation (\ref{eq:fidelspin}), obtained
from a Monte Carlo thermal average of equation (\ref{eq:fidelspinunereal}) over 4000 realisations and temporal maximisation around the cat-state time
(circles with error bars) or peak fidelity  {\rien ${\cal F}_{\rm spin}^{\rm Bog }$} from the Bogoliubov approximation (\ref{eq:fidelspinbogo}) at the cat-state time such that $A(t)=\pi/2$ (upper solid line).
{\rien Dashed red line: lower bound ${\cal F}_{\rm spin}^{\rm Bog, minor}$ on ${\cal F}_{\rm spin}^{\rm Bog}$, as given by equation (\ref{eq:minorant}).
Contrarily to $\mathcal{F}$ and to ${\cal F}_{\rm spin}^{\rm Bog, minor}$, which are universal functions of $k_BT/\Delta$,} $\mathcal{F}_{\rm spin}$ and ${\cal F}_{\rm spin}^{\rm Bog }$ depend on the particle number $N$ and on the interaction strength, adiabatically ramped up and down between
$0$ and the $a$-$a$ and $b$-$b$ scattering length $a_{\rm f}$, see subsection \ref{subsec:fidelspin}. Here the parameters are the ones of figure \ref{fig:g1T}:
$N=300$, $4\pi a_{\rm f}/L=0.0667$ and $t_{\rm ramp}=20 t_{\rm ramp}^{\rm adiab}$.}
The temperature is expressed in units of $\Delta/k_B$, where $\Delta=\frac{\hbar^2 (2\pi)^2}{2m L^2}$ is the minimal excitation energy.}
\label{fig:fidel}
\end{figure}

In practice, the system is prepared at a nonzero temperature $T$. The fidelity of the cat-state preparation, less
than one, can be obtained by the following general reasoning. Let us call $\hat{U}$ the unitary evolution operator during $t_{\rm cat}$
mapping the initial zero-temperature system state $|\psi_0(0^-)\rangle$ onto the cat state $|\rm{cat}\rangle$ (with fidelity one):
\be
\hat{U} |\psi_0(0^-)\rangle = |\rm{cat}\rangle \,.
\ee
If the system is prepared in an initial state $|\psi(0^-)\rangle$ orthogonal to $|\psi_0(0^-)\rangle$, for example in an excited
eigenstate, then the state produced at
time $t_{\rm cat}$ by the same preparation procedure will be orthogonal to the target state $|\rm{cat}\rangle$, {\rien which} corresponds
to a zero fidelity. If the system is prepared in the density operator $\hat{\rho}$, the cat state is obtained with a fidelity
\be
\mathcal{F}=\langle\psi_0(0^-)|\hat{\rho}|\psi_0(0^-)\rangle =P_0
\ee
where $P_0$ is the probability that the system is initially in its ground state. In practice, $\hat{\rho}$ corresponds to
the canonical ensemble at temperature $T$ for an ideal gas in internal state $|a\rangle$. If $T$ is small enough as compared
to the critical temperature $T_c$, one can consider that the condensate is never empty and one can relax the condition that
the number of {\rien noncondensed} particles is less than or equal to the total particle number \cite{Navez,Cartago}. In a given single-particle
mode of wavevector $\kk$, the number of excitations $n_\kk$ then follows the usual exponential law
\be
P_\kk(n_\kk=n) = [1-\exp(-\beta E_k)] e^{-\beta E_k n}, \ \ \ n\in \mathbb{N}
\ee
with $\beta=1/(k_B T)$, $E_k=\hbar^2 k^2/(2m)$ and $\kk\in \frac{2\pi}{L} \mathbb{Z}^{3*}$. The system is in its ground state
if all modes are {\rien in their ground state}. This leads to the cat-state fidelity
\be
\mathcal{F} = \prod_{\kk\neq\OO} (1-e^{-\beta E_k}) \,.
\label{eq:fidelcan}
\ee
This result {\rien is} a universal function of $k_B T/\Delta$, where $\Delta=E_{2\pi/L}=\hbar^2(2\pi/L)^2/(2m)$ 
is the minimal excitation energy, that is the energy gap. It is plotted {\rien as a black solid line} in Fig.~\ref{fig:fidel}. 
This shows that one must have initially a small number of excitations in the system in order to have a fidelity
close to one, hence the stringent requirement on temperature:
\be
k_B T < \frac{\Delta}{4} = \frac{\hbar^2 (2\pi)^2}{8m L^2}
\label{eq:condseveretemp}
\ee
We take for the box size $L=1\,\mu$m  to have about the same chemical potential as in the harmonic trap, for the parameters of Fig.~\ref{fig:experiment}. The energy $\Delta/4$ then corresponds to a temperature of 28 nK. {\rien This temperature is close to the range $\approx 40-30$ nK already accessed by direct in situ evaporative cooling \cite{Dalibard40nK,Allard30nK}. It might also} be reached by using as a coolant the subnanokelvin gases prepared in very weak traps \cite{KetterlenK}.

\subsection{Contrast at nonzero temperature}
\label{subsec:contraste}

We now calculate the $g^{(1)}(t)$ function {\rien of the condensate} in the multimode case (see endnote \cite{endnote55})
within the Bogoliubov approximation, in its $U(1)$-symmetry preserving
version \cite{CastinDum,Gardiner}. The number-conserving noncondensed fields $\hat{\Lambda}_\sigma(\rr,t)$ {\rien in spin state $\sigma$} can be generally expanded over Bogoliubov modes, which
are here plane waves. Due to $g_{ab}=0$, the two spin components decouple.  Due to the adiabatic interaction ramp, 
the mode amplitudes are, up to a global phase factor, given by the instantaneous Bogoliubov steady state expressions.
After the $\pi/2$ pulse, we thus get
\begin{multline}
\left(\begin{array}{c}\hat{\Lambda}_\sigma(\rr,t) \\ \hat{\Lambda}^\dagger_\sigma(\rr,t)\end{array}\right)_{\rm adiab}
=\\
\sum_{\kk\neq\OO} \Big[ \hat{b}_{\sigma,\kk} (0^+) e^{-i\int_0^t d\tau \epsilon_k(\tau)/\hbar} 
\left(\begin{array}{c}U_k(t) \\ V_k(t)\end{array} \right) \frac{e^{i\kk\cdot\rr}}{V^{1/2}}  \\
+\hat{b}^\dagger_{\sigma,\kk} (0^+) e^{i\int_0^t d\tau \epsilon_k(\tau)/\hbar}
\left(\begin{array}{c} V_k(t) \\ U_k(t)\end{array} \right) \frac{e^{-i\kk\cdot\rr}}{V^{1/2}}\Big]
\label{eq:devCD}
\end{multline}
with the instantaneous real amplitudes and {\rien energies}
\bea
U_k(t)+V_k(t) &=& \frac{1}{U_k(t)-V_k(t)} =\left(\frac{E_k}{E_k+2\mu_\sigma(t)}\right)^{1/4} \label{eq:modinst} \\
\epsilon_k(t) &=& [E_k(E_k+2\mu_\sigma(t))]^{1/2} \,. \label{eq:enerBoginst}
\eea
Here the instantaneous chemical potential $\mu_\sigma(t)$ in internal state $\sigma$ is given in the mean-field approximation
$\mu_\sigma(t)=g(t)\bar{N}_\sigma/V$ where $\bar{N}_\sigma=N/2$ is the mean number of particles 
{\rien in that spin state}.
The quasi-particle annihilation and creation operators $\hat{b}_{\sigma,\kk}$ and $\hat{b}^\dagger_{\sigma,\kk}$ 
obey the usual bosonic commutation relations {\rien at equal times}. The operators {\rien for the} numbers of quasi-particles
$\hat{n}_{\sigma,\kk}{\rien =\hat{b}^\dagger_{\sigma,\kk}\hat{b}_{\sigma,\kk}}$ 
are constants of motion in the Bogoliubov approximation, which neglects the quasi-particle 
interactions, {\rien and coincide here with the particle number operators at time $0^+$ since
the gas is still non interacting immediately after the pulse:
\be
\hat{n}_{\sigma,\kk}(0^+)= (\hat{c}_{\sigma,\kk}^\dagger\hat{c}_{\sigma,\kk})(0^+) \ \ \forall \kk\neq\OO \,.
\label{eq:nqpestnp}
\ee
}

The first-order coherence function {\rien of the condensate} in the multimode case is defined similarly to {\rien equation} (\ref{eq:g1}) as
\be
g^{(1)}(t) = \frac{\langle \hat{c}^\dagger_{a,\OO}(t) \hat{c}_{b,\OO}(t)\rangle}
{\langle \hat{c}^\dagger_{a,\OO}(0^+) \hat{c}_{b,\OO}(0^+)\rangle} \,.
\label{eq:defg1multi}
\ee
We use the
usual modulus-phase representation $\hat{c}_{\sigma,\OO}=\exp(i\hat{\theta}_\sigma) [\hat{c}^\dagger_{\sigma,\OO}\hat{c}_{\sigma,\OO}]^{1/2}$
where $\hat{\theta}_\sigma$ is the condensate phase operator in spin state $\sigma$, 
{\rien canonically conjugated to the operator number of particles in the condensate mode of that spin state,
$[\hat{\theta}_\sigma,\hat{c}_{\sigma,\OO}^\dagger\hat{c}_{\sigma,\OO}]=-i$,}
and we perform the usual approximation replacing
the weakly fluctuating moduli by constants, so that
\be
g^{(1)}(t) \simeq \langle e^{-i[\hat{\theta}_a(t)-\hat{\theta}_b(t)]}\rangle \,,
\label{eq:g1dth}
\ee
which expresses the fact that the loss of contrast is due to the condensate phase {\rien spreading} dynamics. At the Bogoliubov order, and neglecting rapidly 
oscillating terms of negligible contribution at long times, the phase evolution {\rien after the
$\pi/2$ pulse} is given by \cite{superdiffusive}
\be
-\hbar \frac{d\hat{\theta}_\sigma{\rien (t)}}{dt} = \mu_0(\hat{N}_\sigma) + \sum_{\kk\neq\OO} \frac{d\epsilon_k}{d N_\sigma}(\hat{N}_\sigma)
\hat{n}_{\sigma,\kk}{\rien (0^+)}
\ee
with $\mu_0(N_\sigma)$ the zero-temperature Bogoliubov chemical potential of a single-component gas with $N_\sigma$ particles \cite{MoraCastin}:
\begin{multline}
\mu_0(N_\sigma) = \frac{g}{V} \left(N_\sigma-\frac{1}{2}\right) \\
+  \frac{g}{V} \sum_{\kk\neq\OO} \left\{V_k(N_\sigma) [U_k(N_\sigma)+V_k(N_\sigma)]
+\frac{N_\sigma g}{2 VE_k}\right\}\\
+\frac{g^2 N_\sigma}{V}\left[\int\frac{d^3k}{(2\pi)^3} \frac{1}{2E_k} -\frac{1}{V} \sum_{\kk\neq \OO}\frac{1}{2E_k}\right] \,.
\label{eq:mu0Bog}
\end{multline}
The last contribution is a finite size effect; the difference between the integral and the sum between square
brackets was evaluated in reference \cite{boite} to be $m C(0)/[(2\pi\hbar)^2L]$ with $C(0)\simeq 8.91364$.
Linearising the dependence of $\mu_0$ and $\frac{d\epsilon_k}{d N_\sigma}$ in $\hat{N}_\sigma$ around $\bar{N}_\sigma$ and integrating over time, we obtain (see endnote \cite{endnote56}) 
\begin{multline}
(\hat{\theta}_a-\hat{\theta}_b)(t)= {\rien (\hat{\theta}_a-\hat{\theta}_b)(0^+)}
-{\rien A(t)}(\hat{N}_a-\hat{N}_b){\rien (0^+)}  \\
-\sum_{\kk\neq\OO} \gamma_k(t) (\hat{n}_{a,\kk}-\hat{n}_{b,\kk}){\rien (0^+)} \,.
\label{eq:dthfin}
\end{multline}
{\rien The time-dependent, dimensionless coefficients are given by}
\be
{\rien A(t)} \!=\!\! {\rien \int_0^t \frac{d\tau}{\hbar}
\Big[\frac{d\mu_0}{d N_\sigma}(\bar{N }_\sigma,\tau)+
\sum_{\kk\neq\OO} \frac{d^2\epsilon_k}{d N_\sigma^2}(\bar{N}_\sigma,\tau)\langle
\hat{n}_{\sigma,\kk}(0^+)\rangle\Big]} \,,\label{eq:coefA} 
\ee
\vspace{-7mm}
\be
\gamma_k(t) = \int_0^t \frac{d\tau}{\hbar} \frac{d\epsilon_k}{d N_\sigma}(\bar{N}_\sigma,\tau) \,.
\label{eq:coefgamk}
\ee
They are affine functions of $t$ for $t>t_{\rm ramp}$. {\rien In particular, one has
\be
A(t) \stackrel{t>t_{\rm ramp}}{=} \chi_T(t-t_{0,T})
\label{eq:Aapres}
\ee
where} the spin nonlinearity coefficient and the retardation time, 
{\rien contrarily to the coefficients $\gamma_k$}, are now temperature dependent (see endnote \cite{endnote57}):
\be
\hbar\chi_T = \frac{d\mu_0}{d N_\sigma}(\bar{N}_\sigma,g=g_{\rm f}) 
+\sum_{\kk\neq\OO} \frac{d^2\epsilon_k}{d N_\sigma^2}(\bar{N}_\sigma,g=g_{\rm f})
\langle\hat{n}_{\sigma,\kk}{\rien (0^+)}\rangle \,.
\ee
To calculate the expectation value in (\ref{eq:g1dth}), {\rien we first exponentiate
the relation (\ref{eq:dthfin}), 
separating the various contributions in three mutually commuting
groups: from left to right, a first group 
containing the operators $\hat{\theta}_a(0^+)$ and $\hat{N}_a(0^+)$, a second one 
containing the quasi-particle number operators $\hat{n}_{\sigma,\kk}(0^+)$ and
a third one containing $\hat{\theta}_b(0^+)$ and $\hat{N}_b(0^+)$.
The Baker-Campbell-Hausdorff formula for two operators $\hat{X}$ and $\hat{Y}$
applied to the first group and to the third group reduces to $\exp(\hat{X}+\hat{Y})=
\exp(\hat{X}) \exp(\hat{Y}) \exp(-\frac{1}{2}[\hat{X},\hat{Y}])$, hence to
\bea
e^{-i\hat{\theta}_a(0^+)+iA(t) \hat{N}_a(0^+)}=e^{-i\hat{\theta}_a(0^+)}
e^{iA(t) \hat{N}_a(0^+)} e^{iA(t)/2}  && \\
e^{-iA(t) \hat{N}_b(0^+)+i\hat{\theta}_b(0^+)}= 
e^{-iA(t)\hat{N}_b(0^+)} e^{i\hat{\theta}_b(0^+)} e^{-iA(t)/2}  &&
\eea
since the commutator of $\hat{\theta}_\sigma(0^+)$ and $\hat{N}_\sigma(0^+)$ is proportional
to the identity.}
It remains 
to express from equations (\ref{eq:apresa}) and (\ref{eq:apresb}) the various
post-pulse operators in terms of the pre-pulse operators:
\bea
(\hat{N}_a-\hat{N}_b){(0^+)} &=& -(\hat{c}^\dagger_{a,\OO}\hat{c}_{b,\OO}+\hat{c}^\dagger_{b,\OO}\hat{c}_{a,\OO})(0^-)
\notag \,, \\
&& + \sum_{\kk\neq\OO} (\hat{n}_{a,\kk}-\hat{n}_{b,\kk}){(0^+)}\\
{(}\hat{n}_{a,\kk}-\hat{n}_{b,\kk}{)}{(0^{+})} &=& -(\hat{c}^\dagger_{a,\kk}\hat{c}_{b,\kk}+\hat{c}^\dagger_{b,\kk}\hat{c}_{a,\kk}){(0^-)} \,, \\
{\rien \hat{\theta}_\sigma(0^+)} &=& {\rien \hat{\theta}_a(0^-) + O(N^{-1/2})} \,,
\eea
using Eq.~(\ref{eq:nqpestnp}) for the first two identities and reference \cite{Casagrande} for the third one,
and to take the expectation value first in the vacuum state of the $\hat{c}_{b,\kk}(0^-)$ ({\rien see the note in the next paragraph})
and then in the thermal state of the  
$\hat{c}_{a,\kk\neq\OO}(0^-)$, eliminating the mode $(a,\kk=\OO)$ using conservation of particle number,
$\hat{c}^\dagger_{a,\OO}\hat{c}_{a,\OO}=
\hat{N}-\sum_{\kk\neq\OO} \hat{c}^\dagger_{a,\kk}\hat{c}_{a,\kk}$. We also need $\langle\hat{n}_{\sigma,\kk}(0^+)\rangle=\frac{1}{2}
[\exp(\beta E_k)-1]^{-1}$ for $\kk\neq\OO$, {\rien and the fact that
$\exp[-i\hat{\theta}_a(0^-)] [\cos A(t)]^{\hat{N}} \exp[i\hat{\theta}_a(0^-)] = [\cos A(t)]^{\hat{N}-1}$.}

{\rien {\it \small Note -- To perform the average on the vacuum in mode $b$, one uses the operatorial relation $\langle 0:b| \exp[i\gamma (\hat{a}^\dagger\hat{b}+\hat{a}\hat{b}^\dagger)]|0:b\rangle=
(\cos\gamma)^{\hat{a}^\dagger\hat{a}}$ where $\hat{a}$, $\hat{b}$ are two bosonic annihilation operators with standard commutation relations,
$\gamma$ is a real number and the expectation value is taken in the vacuum state of $\hat{b}$. As $\hat{a}^\dagger\hat{b}+\hat{a}\hat{b}^\dagger$
conserves the total boson number, it suffices to prove the relation in a Fock state $|n_a:a\rangle$, that is to evaluate 
$\langle S_z=n_a/2|\exp(2i\gamma \hat{S}_x)|S_z=n_a/2\rangle$ according to equation (\ref{eq:spintotal}). Up to a rotation of angle $\pi/2$ around $Oy$, 
this is also $\langle S_x=n_a/2|\exp(-2i\gamma \hat{S}_z)|S_x=n_a/2\rangle={}_{n_a}\langle \frac{\pi}{2};0|\exp(-2i\gamma \hat{S}_z)|\frac{\pi}{2};0\rangle_{n_a}$.
The sought relation then results from the known property $\exp(-2i\gamma \hat{S}_z)|\frac{\pi}{2}; \varphi\rangle_{n_a}=|\frac{\pi}{2};\varphi-2\gamma\rangle_{n_a}$
of the phase states. -- }}

\begin{figure}[htb]
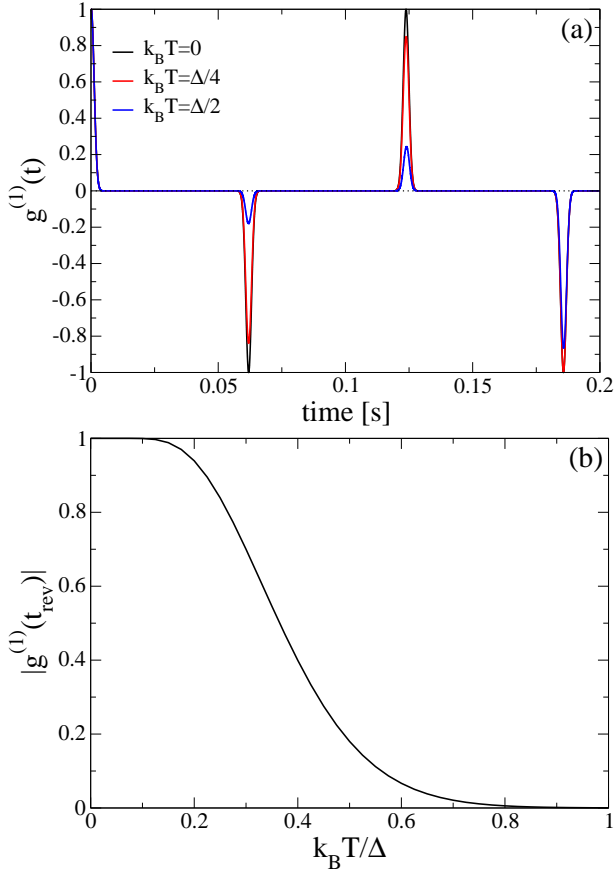

\begin{center}
\includegraphics[width=0.45\textwidth,clip=]{Fig7a.eps}
\includegraphics[width=0.45\textwidth,clip=]{Fig7b.eps}
\end{center}
\caption{First order coherence function $g^{(1)}(t)$ {\rien of the condensate, as given by equation} (\ref{eq:defg1multi}) in the multimode Bogoliubov approximation (\ref{eq:g1bogfin}).
{\rien $N=300$ lossless} ${}^{87}$Rb atoms {\rien are} initially prepared in internal state {\rien $|a\rangle=|F=1,m_F=-1\rangle$} in a box $[0,L]^3$ {\rien with periodic boundary conditions}, $L=1\, \mu$m, at temperature $T$ in the absence of interactions.
{\rien At $t=0^+$ they are} 
subjected to a $\pi/2$-pulse towards the internal state {\rien $|b\rangle=|F=1, m_F=1\rangle$} and to an adiabatic Hann ramping (\ref{eq:Hann}) 
of the interaction strength with a duration $t_{\rm ramp}=20\, t_{\rm ramp}^{\rm adiab}=0.21$ ms, where $t_{\rm ramp}^{\rm adiab}$ is defined in (\ref{eq:condadiabtramp}), up to the final value $g_{\rm f}=4\pi\hbar^2 a_{\rm f}/m$, with
the $a$-$a$ and $b$-$b$ scattering {\rien lengths $a_{\rm f}=100.4 \, a_0$}. In (a): {\rien $g^{(1)}$} as a function of time
for temperatures $k_BT=0$ {\rien (black solid line)}, $k_BT=\Delta/4$ {\rien (red solid line)}, {\rien $k_BT=\Delta/2$ (blue solid line),  from bottom to top at the first {\rien and third} revival times {\rien and from top to bottom at the second revival time}}. In (b): {\rien its absolute value} as a function of temperature at the first revival time {\rien (at which $|g^{(1)}(t)|$ has a maximum)}. {\rien We recall that $\Delta=\frac{\hbar^2(2\pi)^2}{2m L^2}$ is the energy of the first single-particle excited state in the box.}
}
\label{fig:g1T}
\end{figure}

We finally obtain the Bogoliubov prediction for the first order coherence function for $N$ bosons prepared at temperature $T$,  {\rien with an interaction ramped up after the $\pi/2$ pulse from its initial value
$0$ to its final value $g_{\rm f}$:}
\be
g^{(1)}(t)\simeq \cos^{N{\rien -1}}[{\rien A(t)}] 
\prod_{\kk\neq\OO} \frac{1-e^{-\beta E_k}}{1-\frac{\cos[\gamma_k(t)+{\rien A(t)}]}
{\cos[{\rien A(t)}]}e^{-\beta E_k}}
\label{eq:g1bogfin}
\ee
{\rien keeping in mind that, after the ramp, that is at times $t>t_{\rm ramp}$, $A(t)=\chi_T(t-t_{0,T})$
as in (\ref{eq:Aapres}).}
In figure \ref{fig:g1T}a, we plot this prediction as a function of time for various temperatures.
As it is apparent from expression (\ref{eq:g1bogfin}), $g^{(1)}(t)$ results for large $N$ from a narrow function $\cos^{N-1}[A(t)]$ selecting thin temporal windows in a slowly varying envelope function.  At low temperature, when only a few noncondensed modes are populated, the envelope function oscillates in time, {\rien which results in a nonmonotonic behavior of the height of the successive revival peaks as one can see in the figure. In particular for $k_BT/\Delta=0.25$ the third revival is almost perfect.
Indeed, as $|\cos^{N-1}[A(t)]|$ is very small for large $N$, except for the $A(t)$ integer multiple of $\pi$, one can to
a good approximation replace $A(t)$ by such an integer multiple in the product over $\kk$ in equation  (\ref{eq:g1bogfin}), 
which results in the envelope function $\mathcal{G}(t)=\prod_{\kk\neq\OO} \frac{1-e^{-\beta E_k}}
{1-\cos[\gamma_k(t)]e^{-\beta E_k}}$. At $k_BT\ll\Delta$, only the ground noncondensed mode degenerate multiplicity
$k=2\pi/L$ is significantly populated, leading to an almost periodic function $\mathcal{G}(t)$ oscillating between
$\simeq 1$ and $\simeq \mathcal{F}^2$ with an angular frequency $\frac{d}{dt}\gamma_{\frac{2\pi}{L}}(t)$, 
where $\mathcal{F}$ is the
cat-state fidelity (\ref{eq:fidelcan}).} In figure \ref{fig:g1T}b,
we show the value of $g^{(1)}(t)$ at the first revival time of the cosine prefactor, $\chi_T(t-t_{0,T})=\pi$, as a function of temperature.
Both plots show that thermal excitations essentially destroy the revival, except at temperatures below the first excited mode energy
$\Delta$.

\subsection{Spin fidelity at nonzero temperature}
\label{subsec:fidelspin}

The fidelity considered in subsection \ref{subsec:fideltot} is {\rien an} orbito-spinorial fidelity: it measures the overlap of the actual physical state
of the system with the target state (\ref{eq:psimulti}), which is an orbito-spinorial cat state. In pratical applications, however, one mainly
measures pure spin observables, that do not act on the orbital part of the many-body state. This is the case for the collective spin operator $\hat{S}_z$
used as a reference observable in the Fisher information of section \ref{sec:HTWROSA}. It is then more appropriate to consider 
a spin fidelity $\mathcal{F}_{\rm spin}$. The main question is whether or not this spin fidelity is significantly less sensitive 
to nonzero temperature effects than the full fidelity $\mathcal{F}$. This question is answered in this subsection.

For simplicity we assume in this subsection that $N$ is an integer multiple of $4$ so that the cat spin state is, according to equation (\ref{eq:cat_even}),
\be
|\mbox{spin cat}\rangle_N=\frac{|+\rangle_N+i |-\rangle_N}{\sqrt{2}}
\label{eq:chatdespin}
\ee
where $|\epsilon\rangle_N$, with $\epsilon=\pm 1$, is the collective spin state with all the $N$ spins in the same state 
$(|a\rangle+\epsilon |b\rangle)/\sqrt{2}$. In a Bose gas with an orbito-spinorial density operator $\hat{\rho}(t)$ at time $t$,
the cat spin state is realized with a spin fidelity
\be
\mathcal{F}_{\rm spin}(t)=\mbox{Tr}\, \left[\hat{\rho}(t) |\mbox{spin cat}\rangle_N {}_N\langle\mbox{spin cat}|\right] \,.
\label{eq:fidelspin}
\ee

We assume that at the initial time $t=0^-$, the Bose gas in a single realisation occupies in the internal state $|a\rangle$ a $N$-boson 
Fock state $|\psi(0^-)\rangle$.
This state samples the ideal gas thermal equilibrium density operator and is characterised by the occupation numbers $(n_\kk)_{\kk\in\frac{2\pi}{L}\mathbb{Z}^3}$
of the single-particle modes of wavevectors $\kk$ in the quantisation volume $[0,L]^3$:
\be
|\psi(0^-)\rangle = \prod_\kk \frac{(\hat{c}_{a,\kk}^\dagger)^{n_\kk}}{(n_\kk!)^{1/2}} |0\rangle
\ee
We use here the Schr\"odinger picture.  After the instantaneous $a\leftrightarrow b$ $\pi/2$ pulse, the state of the Bose gas is, 
according to equations (\ref{eq:apresa},\ref{eq:apresb}),
\begin{multline}
|\psi(0^+)\rangle=\prod_\kk \frac{\left(\frac{\hat{c}^\dagger_{a,\kk}+\hat{c}^\dagger_{b,\kk}}{\sqrt{2}}\right)^{n_\kk}}{(n_\kk!)^{1/2}}|0\rangle \\
= \sum_{(n_{a,\kk})_\kk} \left[\prod_\kk P_{n_\kk}(n_{a,\kk})\right]^{1/2} |a:(n_{a,\kk})_\kk,b:(n_{b,\kk})_\kk\rangle 
\end{multline}
In the second form, obtained from the first one by using the binomial theorem, the ket is the Fock state with mode occupation numbers 
$n_{a,\kk}$ in internal state $a$ and $n_{b,\kk}$ in internal state $b$, $P_n(n_a)=\frac{n!2^{-n}}{n_a!n_b!}$ is the classical binomial
probability that $n$ incoming particles are split into $n_a$ particles in the output channel $a$ and
$n_b=n-n_a$ in the output channel $b$, one sets $n_{b,\kk}=n_\kk-n_{a,\kk}$ and the sum runs over all the occupation numbers
$(n_{a,\kk})_\kk$ such that $0\leq n_{a,\kk} \leq n_\kk$.

The system then evolves as follows during the time $t$. One switches on adiabatically the interaction strength $g_{aa}=g_{bb}$ 
from $0$ to $g_{\rm f}$ during $t_{\rm ramp}$ with the Hann half-ramp (\ref{eq:enerBoginst}) (we recall that $g_{ab}=0$ at all times). The 
interaction strength remains constant during the time $t-2 t_{\rm ramp}$. It is then switched off adiabatically over
the time interval $[t-t_{\rm ramp},t]$ with the time-reversed Hann half-ramp. In this process, the ideal gas Fock state 
$|a:(n_{a,\kk})_\kk,b:(n_{b,\kk})_\kk\rangle$ is adiabatically turned into a Fock state of Bogoliubov quasi-particles, with the same
occupation numbers for $\kk\neq\mathbf{0}$ (see endnote \cite{endnote58}), 
and is turned back into itself when the interactions are switched off, up to a global phase shift given by
the time integral of the instantaneous eigenenergy $E$ divided by $\hbar$, with
\begin{multline}
E((n_{a,\kk})_\kk,(n_{b,\kk})_\kk,t)= \\
\sum_{\sigma=a,b} \left[E_0(N_\sigma,t) + \sum_{\kk\neq\mathbf{0}}\epsilon_\kk(N_\sigma,t)
n_{\sigma,\kk}\right]
\label{eq:energie}
\end{multline}
The Bogoliubov eigenenergies $\epsilon_\kk$ as functions of the total number of particles $N_\sigma=\sum_\kk n_{\sigma,\kk}$ in internal state
$\sigma$ are given by equation (\ref{eq:enerBoginst}). The ground-state energy $E_0(N_\sigma)$ of $N_\sigma$ interacting bosons 
in internal state $\sigma$ is obtained by integrating $\mu_0(N_\sigma)$ over $N_{\sigma}$ in equation (\ref{eq:mu0Bog})
(knowing that $E_0(N_\sigma=0)=0$). So at time $t$ the Bose gas is in the state
\begin{multline}
|\psi(t)\rangle = \sum_{(n_{a,\kk})_\kk} \left[\prod_\kk P_{n_\kk}(n_{a,\kk})\right]^{1/2}  \\
\times e^{-i\int_0^t d\tau E((n_{a,\kk})_\kk,(n_{b,\kk})_\kk,\tau)/\hbar} |a:(n_{a,\kk})_\kk,b:(n_{b,\kk})_\kk\rangle
\label{eq:psiunereal}
\end{multline}
As shown in Appendix \ref{appen:fidelspin}, the spin fidelity corresponding to the single-realisation
density operator $\hat{\rho}(t)=|\psi(t)\rangle\langle\psi(t)|$ is then
\begin{multline}
\mathcal{F}_{\rm spin}^{\rm single}(t) = 
\left|\sum_{(n_{a,\kk})_\kk} \frac{1-i(-1)^{S_z}}{\sqrt{2}} \left[\prod_\kk 
P_{n_\kk}(n_{a,\kk})\right]\right. \\
\left.
\phantom{\left[\prod_\kk P_{n_\kk}(n_{a,\kk})\right]}
\times e^{-i\int_0^t d\tau E((n_{a,\kk})_\kk,(n_{b,\kk})_\kk,\tau)/\hbar}\right|^2
\label{eq:fidelspinunereal}
\end{multline}
where $S_z=(N_a-N_b)/2$ (see endnote \cite{endnote59}).
It remains to average this result over the thermal canonical distribution of the $(n_\kk)_\kk$ in the initial 
ideal Bose gas to obtain the sought spin fidelity,
\be
\mathcal{F}_{\rm spin}(t) = \sum_{(n_\kk)_{\kk\neq\mathbf{0}}} \mathcal{F}_{\rm spin}^{\rm single}(t) \prod_{\kk\neq\mathbf{0}}
(1-e^{-\beta E_k}) e^{-\beta E_k n_\kk}
\ee
where $E_k=\frac{\hbar^2k^2}{2m}$ and the number of condensate particles $n_{\mathbf{0}}$ is adjusted in each realisation to have a
fixed total number $N$ of particles, $n_{\mathbf{0}}=N-\sum_{\kk\neq\mathbf{0}} n_\kk$.
This average can in practice be taken with a Monte Carlo simulation, and the local maximum of $\mathcal{F}_{\rm spin}(t)$ close to the expected 
cat-state formation time $t_{\rm cat}$ can be found numerically. The resulting spin fidelity for the physical parameters of Fig.~\ref{fig:g1T}
is plotted as symbols with error bars in Fig.~\ref{fig:fidel}, as a function of temperature. 
As expected, it is larger than the orbito-spinorial fidelity $\mathcal{F}$, plotted as a black (lower) solid line in that figure. 
Unfortunately, over the temperature range where it is larger than $1/2$, the spin fidelity is only slightly larger than the orbito-spinorial fidelity $\mathcal{F}$. 
This means that the stringent temperature requirement (\ref{eq:condseveretemp}) also applies to the spin cat-state formation.

We now go through a sequence of approximations to get a more inspiring analytical result and some physical explanation of the sensitivity
of $\mathcal{F}_{\rm spin}$ to temperature. First, as we did in section \ref{subsec:contraste}, we take advantage of the fact that, in the large
$N$ limit, $N_\sigma$ has weak relative fluctuations around its mean value $\bar{N}_\sigma=N/2$, $(N_\sigma-\bar{N}_\sigma)/N\approx N^{-1/2}$.
Expanding the energy $E((n_{a,\kk})_\kk,(n_{b,\kk})_\kk,t)$ in equation (\ref{eq:energie}) to second order in $N_\sigma-\bar{N}_\sigma$ and replacing the coefficients
of the quadratic terms by their thermal averages we obtain
\begin{multline}
\int_0^t d\tau E((n_{a,\kk})_\kk,(n_{b,\kk})_\kk,t)/\hbar \simeq  \\
\int_0^t d\tau [2E_0(\bar{N}_\sigma,\tau)+\sum_{\kk\neq\mathbf{0}} \epsilon_k(\bar{N}_\sigma,\tau) n_\kk]/\hbar \\
+\left[\sum_{\kk\neq\mathbf{0}} \gamma_k(t) (n_{a,\kk}-n_{b,\kk})\right]S_z +A(t) S_z^2
\label{eq:Hquad}
\end{multline}
The first contribution {\rien in equation (\ref{eq:Hquad})} does not depend on $S_z$ nor on the $(n_{a,\kk})_{\kk\neq\mathbf{0}}$ and it will not contribute at all to the spin fidelity.
In the other contributions, the time-dependent coefficients $\gamma_k(t)$ and $A(t)$ are given by equations (\ref{eq:coefA},\ref{eq:coefgamk}).
Second, in the spirit of the particle-number-conserving Bogoliubov methods \cite{CastinDum,Gardiner}, we take as independent variables in each internal state
$\sigma$ the total number of particles $N_\sigma$ and the occupation numbers $(n_{\sigma,\kk})_{\kk\neq\mathbf{0}}$ of the Bogoliubov modes. This is here
an approximation as the $\pi/2$ pulse introduces a small correlation between the difference of the total particle numbers $N_a-N_b$  and the difference
of the noncondensed particle numbers $N_{a,\mathrm{nc}}-N_{b,\mathrm{nc}}$, of the order of $f_{\rm nc}^{1/2}$, where $f_{\rm nc}$ is the initial noncondensed fraction 
(see endnote \cite{endnote60}).
In practice, in expression (\ref{eq:fidelspinunereal}), we perform to leading order in $f_{\rm nc}$ the substitution
\be
\sum_{n_{a,\mathbf{0}}=0}^{n_\mathbf{0}} P_{n_\mathbf{0}}(n_{a,\mathbf{0}}) \rightarrow \sum_{N_a=0}^{N} P_N(N_a)
\label{eq:substitution}
\ee
Finally, using the identity 
\be
\frac{1-i(-1)^{S_z}}{\sqrt{2}} = e^{-i\frac{\pi}{4}} e^{i\frac{\pi}{2} S_z^2} 
\ee
valid for $N$ integer multiple of $4$, we obtain the Bogoliubov approximation for the single realisation spin fidelity
\begin{multline}
\mathcal{F}^{\rm single, Bog}_{\rm spin}(t)= \left|\sum_{N_a=0}^N P_N(N_a) e^{iS_z^2[\frac{\pi}{2}-A(t)]} 
\phantom{\left[\prod_{\kk\neq\mathbf{0}} P_{n_\kk}(n_{a,\kk}) e^{-iS_z\gamma_k(t)(n_{a,\kk}-n_{b,\kk})}\right]}\right.\\
\left. \times \sum_{(n_{a,\kk})_{\kk\neq\mathbf{0}}} \left[\prod_{\kk\neq\mathbf{0}} P_{n_\kk}(n_{a,\kk})  
e^{-iS_z\gamma_k(t)(n_{a,\kk}-n_{b,\kk})}\right]\right|^2
\label{eq:FspinBogunereal}
\end{multline}
The sum over each $n_{a,\kk}$ from $0$ to $n_\kk$ can be evaluated analytically with the binomial theorem. The modulus square of the result can be averaged analytically
over the thermal distribution of the $(n_\kk)_{\kk\neq\mathbf{0}}$ using
$\langle \alpha^{n_\kk}\rangle=[1+\bar{n}_k(1-\alpha)]^{-1}$ here with $|\alpha|\leq 1$, which gives
\begin{multline}
\mathcal{F}_{\rm spin}^{\rm Bog}(t) = \\
\sum_{N_a,N'_a=0}^{N} \frac{P_N(N_a) P_N(N'_a)\cos[(S_z^2-{S'_z}^2)(A(t)-\frac{\pi}{2})]}
{\prod_{\kk\neq\mathbf{0}}\{1+\bar{n}_k [1-\cos(\gamma_k(t)S_z)\cos(\gamma_k(t)S'_z)]\}}
\label{eq:fidelspinbogo}
\end{multline}
where $\bar{n}_k=(e^{\beta E_k}-1)^{-1}$ is the initial mean occupation number of the mode $\kk$ in the internal state $a$.
The approximate result (\ref{eq:fidelspinbogo}) is readily evaluated numerically as a function of time
for the physical parameters of Fig.~\ref{fig:g1T}. It is found that the sought
spin fidelity peak is located extremely close to the expected cat-state formation time $t_{\rm cat}$, such that $A(t_{\rm cat})=\frac{\pi}{2}$,
and its value, plotted as a red (upper) solid line in Fig.~\ref{fig:fidel}, is in very good agreement with the Monte Carlo results
(red circles) resulting from the full expression (\ref{eq:fidelspinunereal}).

{\rien {\it \small Note -- The approximation made in (\ref{eq:Hquad}), consisting of the replacement of coefficients
of the quadratic terms by their thermal averages, is not essential. Without it, one obtains
\[
\mathcal{F}_{\rm spin}^{\rm Bog}(t) = \!\!\!\!
\sum_{N_a,N'_a=0}^{N}\!\!\! \frac{P_N(N_a) P_N(N'_a)e^{-i (S_z^2-S_z'^2)[A_0(t)-\frac{\pi}{2}]}}
{\prod_{\kk\neq\mathbf{0}}[1+\bar{n}_k (1-D_k)]} 
\]
where $A_0(t)$ is the zero temperature value of $A(t)$, 
$D_k=e^{-i\alpha_k(S_z^2-S_z'^2)}\cos(\gamma_k(t)S_z)\cos(\gamma_k(t)S'_z)$ and 
$\alpha_k(t)=\frac{1}{2}\int_0^t \frac{d\tau}{\hbar} \partial_{N_\sigma}^2
\epsilon_k (\bar{N}_\sigma,\tau)$\,. We have verified that this result is very close to the less refined approximation  
(\ref{eq:fidelspinbogo}) for the parameters of Fig.~\ref{fig:fidel}. -- }}

A physical insight in the temperature sensitivity of the spin fidelity is obtained 
by rewriting the single realisation spin fidelity at the cat-state time $t_{\rm cat}$ from equation (\ref{eq:FspinBogunereal}) as
\be
\mathcal{F}^{\rm single, Bog}_{\rm spin}(t_{\rm cat})= \left|\left\langle \cos^N \frac{\Delta \theta_{\rm th}}{2}\right\rangle_{\rm partition}\right|^2
\label{eq:lumiere}
\ee
where the average is taken over the partition noise {\rien in the noncondensed modes} accompanying the $\pi/2$ pulse, that is with the binomial probability distribution $P_{n_\kk}(n_{a,\kk})$
for each $n_{a,\kk}$, and 
\be
\Delta \theta_{\rm th}=\sum_{\kk\neq\mathbf{0}} \gamma_k(t_{\rm cat}) (n_{a,\kk}-n_{b,\kk})
\label{eq:deltatheta}
\ee
is a random thermal shift of the $a$-$b$ condensate relative phase, already present in operatorial form in equation (\ref{eq:dthfin}).
The form (\ref{eq:lumiere}) is obtained by summing over $N_a$ in equation (\ref{eq:FspinBogunereal}), {\rien taking into account the fact that $A(t_{\rm cat})=\pi/2$}.
This is exactly the single realisation spin fidelity that one would obtain if the collective spin was in the quantum state (see endnote \cite{endnote61})
\be
|\psi_{\rm spin}^{\rm single}(t_{\rm cat})\rangle = \langle \: e^{i\Delta\theta_{\rm th}\hat{S}_z}|\mbox{spin cat}\rangle_N \: \rangle_{\rm partition}
\label{eq:suggestive}
\ee
that is in a coherent superposition of rotated cat states (what appears here is the operator $\hat{S}_z$).
As the cat-state time scales as $N$, the coefficients $\gamma_k(t_{\rm cat})$ scale as $N^0$ and are of order unity. This shows that the presence of a single
thermal excitation in the initial state of the system, by activating the partition noise, will give quantum fluctuations of $\Delta \theta_{\rm th}$ of order unity,
sufficient to compromise the cat-state fidelity (see endnote \cite{endnote62}). 
This high sensitivity to thermal excitations was anticipated in reference \cite{Mimoun}.

{\rien Equation (\ref{eq:lumiere}) is not only physically appealing, it also provides a lower bound to 
the peak spin fidelity ${\cal F}_{\rm spin}^{\rm Bog}(t_{\rm cat})$ that is very close to the actual value for large $N$.
Indeed, when $N\gg1$ in a fixed volume and at a fixed temperature, $\cos^N(\Delta \theta_{\rm th}/2)$ is a very narrow function of $\Delta \theta_{\rm th}$ with a width smaller than the discreteness of the distribution of  $\Delta \theta_{\rm th}$. As a consequence, only the realisations with $\Delta \theta_{\rm th}=0$ contribute significantly. As the coefficients $\gamma_k$ for different wave numbers $k$ are in general incommensurable, this imposes that for all allowed wave numbers $k\neq 0$ in the quantization volume:
\be
\sum_{\kk \, /\,  ||\kk||=k} (n_{a,\kk} -n_{b,\kk}) = 0 \,.
\ee
For a given realisation of the initial thermal occupation numbers $n_\kk$,  this occurs for a given $k$ with the binomial probability 
$P_{N_k}(N_k/2)$ where $N_k$ is the total number of initial thermal excitation in the degenerate manifold of wave number $k$,
$N_k=\sum_{\kk \, /\,  ||\kk||=k} n_\kk$. Note that this probability is zero for odd $N_k$. As $N$ is here even, $\cos^N$ is nonnegative and we obtain after thermal average the inequality $
{\cal F}_{\rm spin}^{\rm Bog }(t_{\rm cat}) \geq  {\cal F}_{\rm spin}^{\rm Bog,\, minor}$
with the lower bound
 \begin{multline}
{\cal F}_{\rm spin}^{\rm Bog,\, minor}   \equiv  \langle \prod_{k \neq 0} [P_{N_k}(N_k/2)]^2 \rangle_{\rm therm}  \\
= {\cal F} \prod_{k\neq 0} {}_3 F_2\left(\frac{1}{2},\frac{d_k}{2},\frac{1+d_k}{2};1,1;e^{-2\beta E_k} \right)
\label{eq:minorant}
\end{multline}
where ${\cal F} $ is the orbito-spinorial fidelity (\ref{eq:fidelcan}), $d_k$ is the degeneracy of the manifold $k$, and ${}_3 F_2$ is a  
generalised hypergeometric function, see \S 9.100 in reference \cite{Gradstein}.
The lower bound (\ref{eq:minorant}) is represented as a dashed red line in Fig.~\ref{fig:fidel}. Remarkably this universal function of $k_BT/\Delta$ is almost indistinguishable from ${\cal F}_{\rm spin}^{\rm Bog }(t_{\rm cat})$ at the scale of the figure} (see endnote \cite{endnote63}).

\section{Conclusion}
\label{sec:conclusion}

We have studied the interaction-induced formation of mesoscopic {\rien quantum} superpositions in bimodal Bose-Einstein condensates including limiting effects such as particle losses and fluctuations of the total number of particles. We have explained how these effects can be compensated, giving two examples for sodium and rubidium {\rien Bose-Einstein condensates}. To quantify the survival of quantum correlations in the presence of decoherence, and their usefulness for metrology, we have calculated the Fisher information and the Wigner function of the obtained state, and we have also shown that, in the presence of losses, there is a simple quantitative relation between the cat-state fidelity and the amplitude of the revival peak in phase contrast. Finally, giving up the two-mode description in a last multimode section, we have described a possible procedure to prepare the initial state, {\rien and we have studied the influence of a nonzero initial temperature on the amplitude of the phase revival and on the cat-state fidelity. Two fidelities are introduced: the full orbito-spinorial fidelity ${\cal F}$ and a purely spinorial fidelity ${\cal F}_{\rm spin}$, defined once the orbital degrees of freedom are traced out}. We find that macroscopic superpositions can be obtained at nonzero temperature {\rien with a high fidelity, with no substantial gain of ${\cal F}_{\rm spin}$ with respect to ${\cal F}$,} provided the temperature is lower than {\rien about one quarter of} the energy of the first single-particle excited state.

\acknowledgments
{\rien We acknowledge useful discussions with Dominique Spehner and Anna Minguzzi. Dominique Spehner made analytical calculations about how to choose parameters to satisfy the compensation condition in the Gaussian regime, that will be exploited in a future work.} During an internship at ENS in 2000, Uffe Poulsen  obtained with a different technique for a one-dimensional harmonically trapped Bose gas in the Hartree-Fock limit similar results for the multimode contrast (\ref{eq:g1bogfin}) (unpublished work). {\rien K. P. acknowledges support from Polish National Science Center Grant No UMO-2014/13/D/ST2/01883.}
{\rien M.F. and P.T. acknowledge financial support from the Swiss National Science Foundation through NCCR QSIT.}

\appendix

{\rien 
\section{Fidelity and revival beyond the constant loss rate approximation
	\label{app:Losses}}
In section \ref{sec:CLRA} we derived a simple relation (\ref{eq:result_Fidelity}) between the cat-state fidelity and the revival amplitude using the constant loss rate approximation. In this appendix we calculate the first correction to this approximation. {\rien As in section \ref{sec:CLRA}, we restrict to the case in which the two components are symmetric and spatially separated.} All analytical results are derived in the frame of the stochastic wavefunction approach, while the numerical results come from the exact diagonalization method described in Appendix A of reference \cite{Krzysiek2}, applied to the master equation.

The initial state is the phase state placed on the equator of a pseudo-Bloch sphere with exactly $N$ atoms, i.e.  $\ket{\psi(t=0)} = \ketph{\theta= \frac{\pi}{2}}{\phi=0 }_N$.

Due to particle losses the state evolves into a mixed state, 
	\begin{equation}
	\hat{\rho}(t) = \sum_{n=0}^N  \hat{\rho}_n (t),
	\label{eq:appA-rho}
	\end{equation}
where $\hat{\rho}_n$ is the unnormalized density matrix corresponding to the restriction of $\hat{\rho}$ to the subspace with exactly $n$ atoms. The trace of the state $\hat{\rho}_n$ is the probability that the total number of atoms is equal to $n$:
	\begin{equation}
	p_n (t) =  \mathrm{Tr} \,\hat{\rho}_n \,.
	\end{equation}
In the stochastic wavefunction approach there is at time $t$ only one stochastic wavefunction with the initial number of atoms, the one that has not experienced any quantum jump:
	\begin{equation}
	\ket{\psi_N(t)} = e^{-i\,t \left[\hat{H} - \frac{i\hbar}{2} \sum_{m,\epsilon} \left(\hat{J}_{\epsilon}^m\right)^{\dagger}\hat{J}_{\epsilon}^m \right] / \hbar}  \ket{\psi(0)} .
	\end{equation}
It means that within this subspace the (unnormalized) state remains pure, i.e.  $\hat{\rho}_N = \ket{\psi_N(t)}\bra{\psi_N(t)}$.

In the lossless case, the total number of atoms is fixed to $N$. Hence the time-dependent fidelity between the state in the lossless case, denoted with $\ket{\psi^{(0)} (t)}$, and  the density matrix  \eqref{eq:appA-rho} depends only on the state restricted to the subspace with the $N$ atoms:
	\begin{eqnarray}
	\mathcal{F}(t) &=&  \mathrm{Tr} \left[ \hat{\rho}(t)\ket{\psi^{(0)} (t)}\bra{\psi^{(0)} (t)} \right] =\left| \braket{\psi_N(t)}{\psi^{(0)} (t)}  \right|^2 \nonumber\\
	 &=& \left|\bra{\psi(0)} e^{ - \frac{t}{2}\sum_{m,\epsilon} \left(\hat{J}_{\epsilon}^m\right)^{\dagger}\hat{J}_{\epsilon}^m}  \ket{\psi(0)}\right|^2
	 	\label{eq:app-fidelity}
	\end{eqnarray}
We  relate the fidelity to the normalized first order correlation function:
	\begin{equation}
	g^{(1)}(t) = \frac{2}{N}\langle\hat{S}_x\rangle(t) = \frac{2}{N}\sum_{n=0}^N \mathrm{Tr} \left[ \hat{S}_x \hat{\rho}_n \right] = 
	\sum_{n=0}^N g^{(1)}_n (t),
		\label{eq:}
	\end{equation}
where $ g^{(1)}_n (t) \equiv\frac{2}{N} \mathrm{Tr} \left[ \hat{S}_x \hat{\rho}_n \right] $ is the contribution to $g^{(1)}(t)$ of the subspace with $n$ atoms.

In what follows we use the notations
$\tilde{\gamma}  \equiv \gamma^{(1)} t_{\rm rev}$ for one-body losses and
$\tilde{\gamma}  \equiv \gamma^{(3)} t_{\rm rev}$ for three-body losses, where $t_{\rm rev}=\pi/\chi$ is the first revival time.
	
\subsection{One-body losses}
	\begin{figure}
	 \includegraphics[scale=0.28]{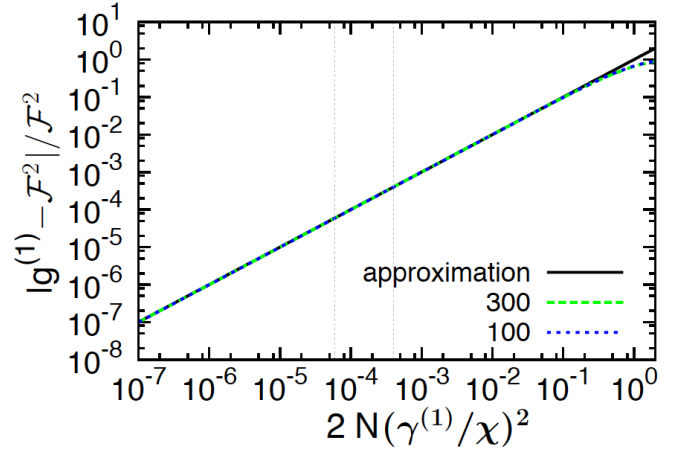}
	\caption{Relative deviations between $g^{(1)}{\rien(t_{\rm rev})}$ and ${\rien {\cal F}^2}{\rien(t_{\rm cat})}$  {\rien in the presence of one-body losses}. {\rien The approximate formula \eqref{eqn:corr1} is represented as a full line, while the dashed lines are exact solutions for $N=100$ and $N=300$. The values of the abscissa $2N (\gamma^{(1)}/\chi)^2$ corresponding to the trapping angular frequency $\omega=2\pi\times 500$ Hz and the scattering length $a=100.4$ Bohr radii and one-body loss rate equal to $K_1=0.01$Hz {\rien are marked  as dotted vertical lines} for $N=100$ (left line) and $N=300$ (right line).
 {\rien Note that as in section \ref{sec:CLRA}, we restrict here to the case in which the two components are symmetric and spatially separated.}	}
	\label{fig:corr1}}
	\end{figure}

We now look at corrections to  the constant loss rate approximation in the presence of one-body losses.
In this case there  are two jump operators: $\hat{J}_{a, 1} = \sqrt{\gamma^{(1)}} \,\hat{a}$ and $\hat{J}_{b, 1} = \sqrt{\gamma^{(1)}} \,\hat{b}$. The fidelity evaluated from Eq. \eqref{eq:app-fidelity} is equal to ${\cal F}(t) =e^{-N\gamma^{(1)} t}$ (exactly as in the constant loss rate approximation).

In the case of one-body losses the full $g^{(1)}$ function at the time $t_{\rm rev}$ can be calculated exactly:
	\begin{equation}
	 g^{(1)} (t_{\rm rev}) = \bb{\frac{(\gamma^{(1)})^2 - \chi^2 e^{-\tilde{\gamma}}}{(\gamma^{(1)})^2 + \chi^2 }}^{N-1}e^{-N \gamma^{(1)} t_{\rm rev}} \,.
	 \label{eq:g1_1body_exact}
	\end{equation}
We quantify the discrepancy between $g^{(1)}(t_{\rm rev})$ and $\left( {\cal F}(t_{\rm cat})\right)^2$  with the relative deviation 
 $|g^{(1)} - {\cal F}^2   | /  {\cal F}^2$, plotted for $N=300$ (green dashed line) and $N=100$ (blue dotted line) in Fig.~\ref{fig:corr1}.

The contribution to $g^{(1)}$ from the subspace with the initial number of atoms reads
	\begin{equation}
	 |g^{(1)}_N (t_{\rm rev})| = \mathcal{F}(t_{\rm cat})^2 = e^{-N \gamma^{(1)} t_{\rm rev}} \,.
	\label{eqn:g1N_1body}
	\end{equation}
Thus, if one restricts to the subspace with $N$ atoms, the fidelity-contrast  relation (\ref{eq:result_Fidelity}) becomes exact. 
The small discrepancy is due to the contributions $g^{(1)}_{n}(t)$ from the other subspaces $n<N$. 
The leading one is 
	\begin{equation}
	g^{(1)}_{N-1} (t_{\rm rev}) = \frac{\bb{N-1}\tilde{\gamma}^2}{\pi^2 + \tilde{\gamma}^2} e^{-\tilde{\gamma} N}(e^{\tilde{\gamma}}+1)
	\approx \frac{2N \tilde{\gamma}^2}{\pi^2}  e^{-\tilde{\gamma} N}\,.
	\end{equation}
By including this correction we obtain the approximate formula
	\begin{multline}
	| |g^{(1)}{\rien(t_{\rm rev})}| -   {\cal F}^2{\rien(t_{\rm cat})}|/ { {\cal F}}^2{\rien(t_{\rm cat})} \approx \bb{\frac{g^{(1)}_{N-1}{\rien(t_{\rm rev})}}{  {\cal F}{\rien(t_{\rm cat})}}}^2 \\ \approx 2N \bb{\frac{\gamma^{(1)}}{\chi}}^2 \,.
	\label{eqn:corr1}
\end{multline}
In Fig.~\ref{fig:corr1} we compare the approximate expression (\ref{eqn:corr1}) and the
exact value of the relative correction calculated from (\ref{eq:g1_1body_exact}).
We note that $g^{(1)}_{N-1}\leq \frac{8}{\bb{\pi e}^2}$, the equality holding for $\gt=\frac{2}{N}$.

\subsection{Three-body losses}
Let us now consider the case of three-body losses.
As the two components do not overlap, there are only two associated jump operators: $\hat{J}_{a, 3} = \sqrt{\gamma^{(3)}} \,\hat{a}^3$ and $\hat{J}_{b, 3} = \sqrt{\gamma^{(3)}} \,\hat{b}^3$. From Eq. \eqref{eq:app-fidelity} we obtain the fidelity 
\begin{eqnarray}
{\cal F}(t) &=& \left|\bra{\psi(0)} e^{ - \frac{\gamma^{(3)} t}{2} \left((\hat{a}^{\dagger})^3\hat{a}^3+(\hat{b}^{\dagger})^3\hat{b}^3\right)}  \ket{\psi(0)}\right|^2 \nonumber\\
&=& \frac{1}{2^N}\left(\sum_{k=0}^N \binom{N}{k} \exp\bb{- h(k)\,\gamma^{(3)} t /2} \right)^2,
\label{eqn:F_3body}
\end{eqnarray}
where $h(k) = \frac{k!}{(k-3)!} + \frac{(N-k)!}{(N-k-3)!} $.

In the case of three-body losses we cannot compute analytically the first order correlation functions.
Using the stochastic wavefunction approach we can however calculate the contributions to $g^{(1)}$ of subspaces with $N$ and $N-3$ atoms:
	\begin{eqnarray}
	g^{(1)}_N (t_{\rm rev}) &=& {\rien (-1)^{N-1}} \frac{1}{2^{N-1}}\sum_{k=0}^{N-1} \binom{N-1}{k} \,, \nonumber \\
	&\times&\exp\bb{-\tilde{\gamma} \bb{h(k) + h(k+1)}} \,,  \label{eqn:g1_3body}\\
	g^{(1)}_{N-3} (t_{\rm rev}) &=&  {\rien (-1)^{N}} \frac{(N-1)!\tilde{\gamma}^2}{2^{N}(N-4)!}\sum_{k=0}^{N-4} \binom{N-4}{k} {\rien K}(k, N)\nonumber \\
	&\times&e^{-\tilde{\gamma} J(k, N)} \,,  \label{eqn:g13_3body}
\end{eqnarray}
where 
\begin{eqnarray}
{\rien K}(k, N) &=& f(k) + f(N-4-k) \,, \\
 f(n) &=& \frac{1+\exp\bb{-\frac{\gt w(n+3)}{2}}}{9\pi^2+\bb{\frac{\gt w(n+3)}{2}}^2} \,,\\
 w(n) &=&174 + 108 n + 18 n^2 - 108 N \nonumber \\
 &-& 36 n N + 18 N^2 \,, \\
 J(n, N) &=& 2 N^3 - 6 n N^2 + 6 n^2 N + 54 n N  - 27 N^2  \nonumber \\ &-& 120 n - 30 n^2 + 121 N -180 \,.
\end{eqnarray}

\begin{figure}
\begin{center}
 \includegraphics[scale=0.28]{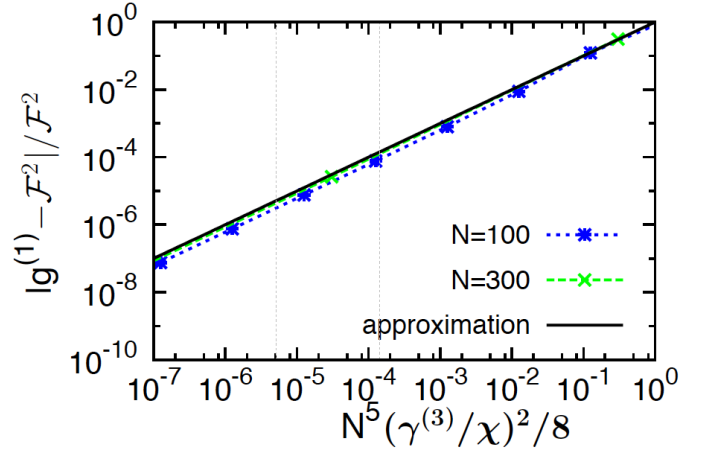}
\caption{Relative deviations between $g^{(1)}{\rien(t_{\rm rev})}$ and {\rien ${\cal F}^2{\rien(t_{\rm cat})}$} {\rien in the presence of three-body losses}. {\rien The approximate formula \eqref{eqn:corr3} is represented as a full line, while the symbols linked by dashed lines are exact solutions for $N=100$ and $N=300$. The values of the abscissa $N^5(\gamma^{(3)}/\chi)^2/8$ corresponding to the trap and loss parameters of Fig.~\ref{fig:experiment}, {\rien are marked as dotted vertical lines} for $N=100$ (left line) and $N=300$ (right line). {\rien Note that as in section \ref{sec:CLRA}, we restrict here to the case in which the two components are symmetric and spatially separated.}}
\label{fig:corr3}}
\end{center} 
\end{figure}

In the limit of large {\rien atom numbers,} the binomial distribution can be approximated with a Gaussian distribution and the sums over $k$  with integrals{\rien :} 
\begin{equation}
 \frac{1}{2^N}\sum_{k=0}^N \binom{N}{k} f(k) \approx \int_{-\infty}^{\infty}dx\, g(x) f(x) \,,
\label{eqn:approx}
\end{equation}
where $g(x) = \frac{1}{\sqrt{\pi N/2}}\exp\bb{-\frac{2 \bb{x-N/2}^2}{N}}$.
Using this continuous limit we approximate Eqs.  \eqref{eqn:F_3body}-\eqref{eqn:g1_3body} with
\begin{eqnarray}
\mathcal{F}({\rien t_{\rm cat}}) &=& \frac{e^{-N (N-2) (N-4) \gt / 8} }{{1+\frac{3}{8}\gt N (N-2) }} \,,\\
g^{(1)}_N (t_{\rm rev})  &=& -\frac{ {\rien (-1)^{N}}e^{-(N-1)(N-2)(N-3) \gt/4}}{\sqrt{1+\frac{3}{2}\gt (N-1)(N-2) }} \,.
\end{eqnarray}
The contribution of {\rien the} subspace with $N-3$ atoms to {\rien $g^{(1)}(t_{\rm rev}) $}, in the limit $\gt N^2 \to 0$, $N \to \infty$ with $\gt N^3$ fixed,  is 
\begin{equation}
g^{(1)}_{N-3} (t_{\rm rev}) \approx \frac{N^5\gt^2}{8\pi^2}\exp\bb{-N^3 \gt /4} .
\end{equation}
{\rien Corrections to the relation (\ref{eq:result_Fidelity}), stating that ${\cal F}(t_{\rm cat}) = |g^{(1)}(t_{\rm rev})|^{1/2}$,  then come from two sources:}
{\rien from the difference} $\mathcal{F}^2 - |g^{(1)}_N|$ and {\rien from the difference} $g^{(1)}  - g^{(1)}_N$.
In the limit of {\rien a} small lost fraction $\gt N^2 \ll 1$ {\rien we obtain}:
\begin{itemize}
 \item[(i)] $|\mathcal{F}^2{\rien(t_{\rm cat})} - |g^{(1)}_N{\rien(t_{\rm rev})}||/g^{(1)}{\rien(t_{\rm rev})} = 
 O(\tilde{\gamma}^2 N^4)$ \,,
 \item[(ii)] $|g^{(1)}{\rien(t_{\rm rev})}  - g^{(1)}_N{\rien(t_{\rm rev})}|/g^{(1)}{\rien(t_{\rm rev})} \approx \frac{1}{8\pi^2} \tilde{\gamma}^2 N^5 = \frac{(\gamma^{\rien (3)})^2 N^5}{8\chi^2}$  \,.
\end{itemize}
The leading corrections come from (ii), as confirmed by Fig.~\ref{fig:corr3}, which compares {\rien the approximate analytical result}
\begin{equation}
|{\rien |g^{(1)}{\rien(t_{\rm rev})}|}  - {\rien {\cal F}}^2{\rien(t_{\rm cat})}|/{\rien {\cal F}}^2{\rien(t_{\rm cat})} \approx \frac{(\gamma^{\rien (3)})^2 N^5}{8\chi^2}
\label{eqn:corr3}
\end{equation}
with an exact numerical calculation.

We note that, both for three-body and one-body losses, {\rien up to a numerical factor $2/(m\pi)^2$}, the relative corrections (\ref{eqn:corr1}) and (\ref{eqn:corr3}) can be interpreted as the product between the number of lost atoms and the fraction of lost atoms. {\rien In the interesting regime in which the number of lost atoms at the revival time is smaller than one (the fidelity and the revival would be killed by the losses otherwise), the corrections are then smaller than the dominant contribution of the losses coming from the $N$ particles subspace, by a factor $1/N$.}
}

\section{Search algorithm}
\label{app:search}
\begin{figure*}[b]
	\begin{center}
		\includegraphics[width=0.9\textwidth]{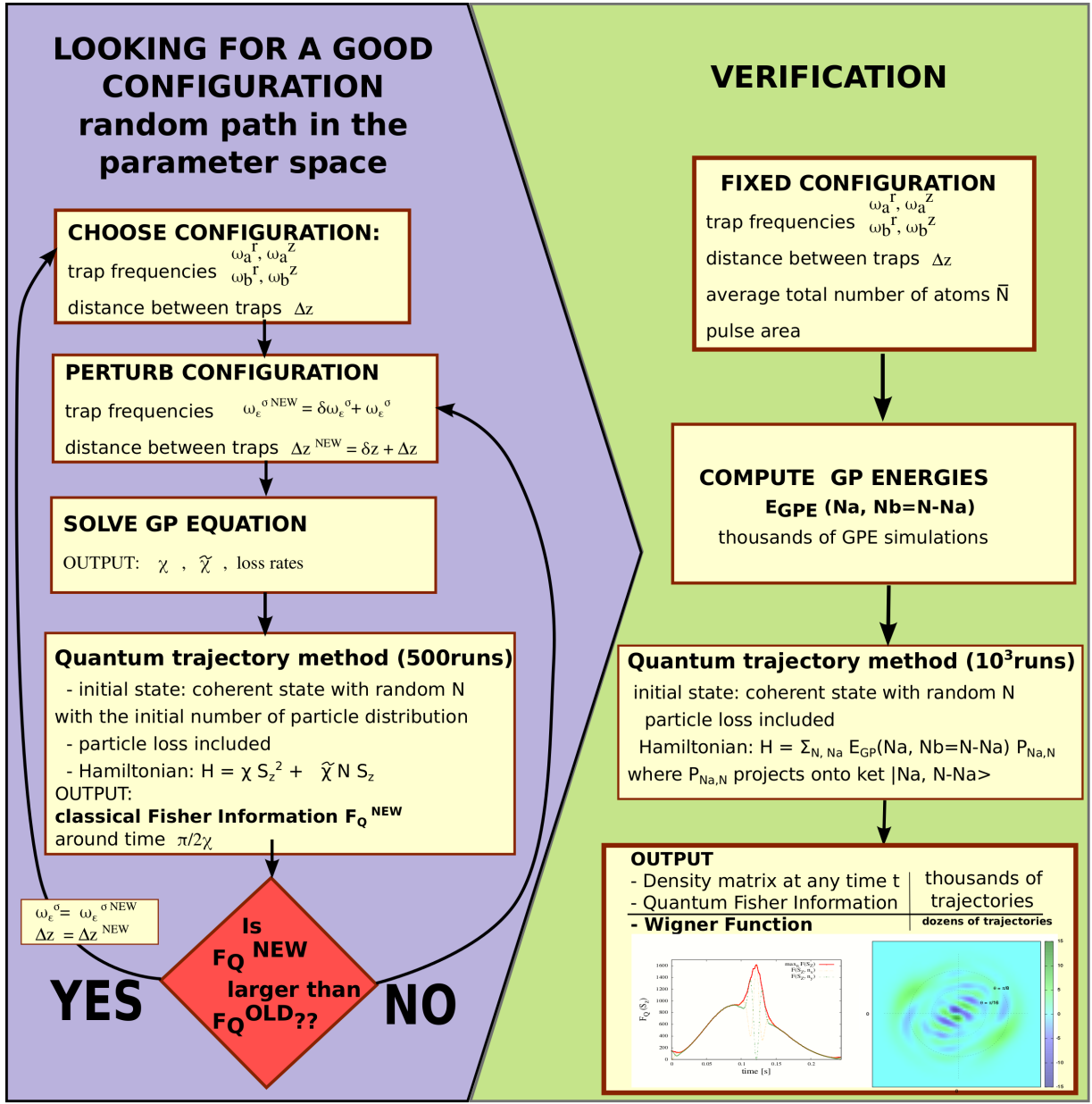}
	\end{center} 
	\caption{
The algorithm searches for good configurations in the parameter space by maximizing the Fisher information and verifies the presence of a {\rien small-amplitude cat} state signaled by high contrast fringes in the Wigner function.} 
	\label{fig:algorithm}
\end{figure*}

\begin{figure}[htb]
	\begin{minipage}[l]{0.48\textwidth}
	\vspace{0.5cm}
	\includegraphics[width=0.80\textwidth]{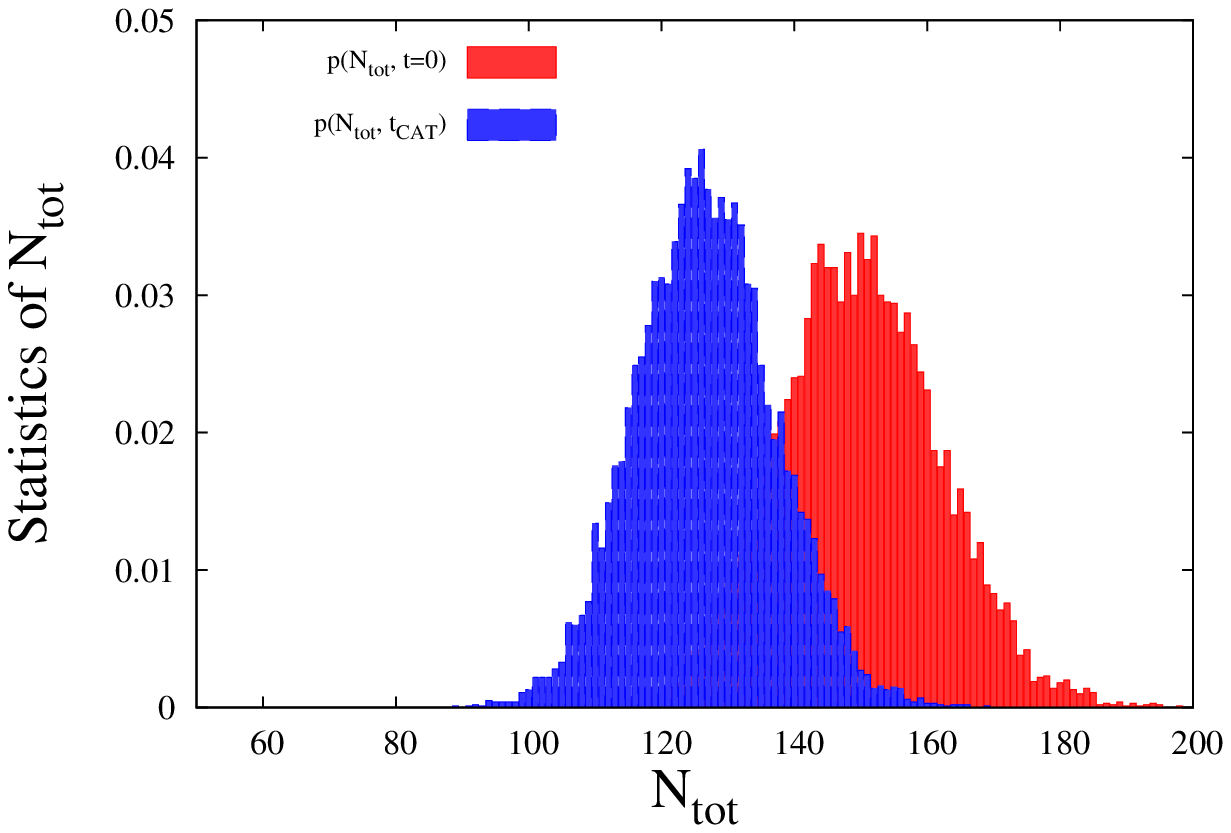}
	\end{minipage}
	\begin{minipage}[l]{0.48\textwidth}
		\begin{tabular}{l|c|c|c|}
		 jump & \#  events & lost in $a$ & lost in $b$  \\ \hline\hline	
		   $a$   &  0.135   & 0.135 &   0 \\
		   $b$   &  3.1   & 0     &   3.1 \\
		   $aa$  &  0.0  & 0.0 &   0.0 \\
		   $ab$  &  0.1   & 0.1 & 0.1 \\
		   $bb$  &  13.0  & 0 &   26.1 \\
		   $aaa$ &  0.0   & 0 &   0.0 
		\end{tabular}
	\end{minipage}	
	\caption{Initial (red {\rien peak on the right}) and the final (blue {\rien peak on the left}) {\rien probability distribution} of the total number of atoms.
	Around 20\% of {\rien the} atoms {\rien are} lost, 
but practically only in the {\rien majority} component $b$. 
	Table: Budget of losses at time $t_{\rm cat}$ . Parameters are the same {\rien as} in {\rien Fig.~\ref{fig:wigS}}.\label{fig:losses_budget}} 
	\end{figure}

\begin{figure}[b]
	\begin{center}
		\includegraphics[width=0.48\textwidth]{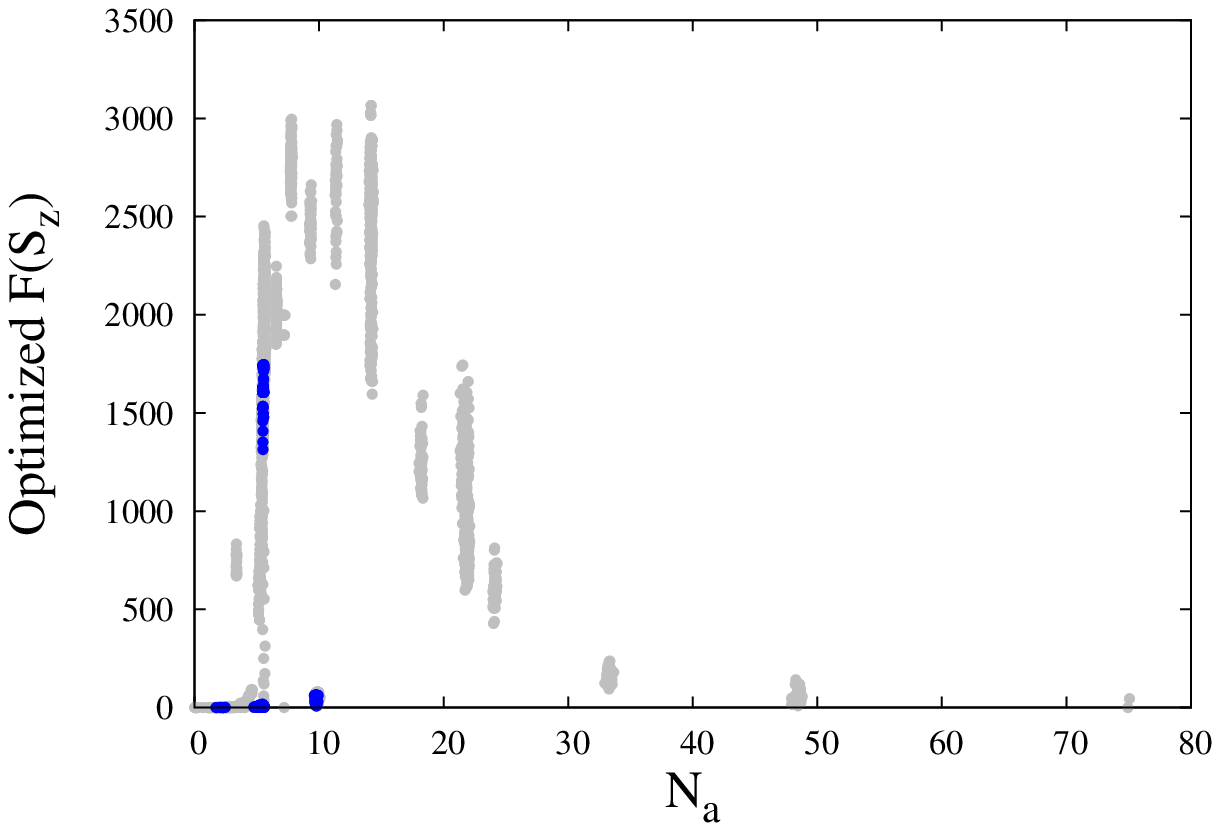}
	\end{center} 
	\caption{
{\rien Example of output of the optimization program described in Appendix \ref{app:search} for the Fisher information at the cat-state time. $N_a$ on the $x$-axis is in the initial mean number of atoms in the minority component. For a given $N_a$ the different points correspond to successive configurations explored by the algorithm in its convergence process.
Restricting to configurations with trap aspect ratios smaller than $20$ we select out the darker points (in blue). Scattering lengths and loss rates are as in Fig.~\ref{fig:wigS} for ${}^{87}$Rb. Fig.\ref{fig:wigS} corresponds to one of the blue points with maximal Fisher information around 1500. 
	\label{fig:fishmax}}}
\end{figure}

The numerical algorithm we use to find optimal parameters within experimental constraints, to create the cat in the case of a hyperfine transition in rubidium or sodium{\rien ,} is described in Fig.~\ref{fig:algorithm}.
We fix the average total number of atoms and the pulse preparing the initial phase state, and {\rien the code varies} {\rien some} parameters to maximize the Fisher Information (\ref{eq:fisher}) at the cat-state time.
{\rien For example, for rubidium,} the variational parameters are the radial trap frequencies $\omega_{\perp}$ assumed to be equal for the two species, the longitudinal frequencies $\omega_{az}$ and $\omega_{bz}$ and the distance between trap centers along $z$, denoted $\Delta z$. 
Once a favorable configuration is found by the algorithm, we proceed to a verification step by calculating 
both the Fisher Information and the Wigner function beyond the $\chi \Sz^2$ approximation, meaning that instead of the Hamiltonian \eqref{eq:H} in the master equation, we use (\ref{eq:hamGPE}).
The Wigner function is defined as
$W(\theta, \varphi) =\sum_{N=0}^{\infty} p(N) W_N(\theta, \varphi)$ where $W_N(\theta, \varphi)$ is normalized to unity \cite{schleich1994}. 
We show a result in figure \ref{fig:wigS} for a particular configuration. {\rien For this particular configuration,} corresponding to the rubidium {\rien 87} case, in Fig.~\ref{fig:losses_budget}
we {\rien show the probability distributions for the total number of atoms and we} summarize the loss budget. 
Note that although $30$ particles are lost on average in the {\rien majority} component, {\rien high} contrast fringes are obtained in the Wigner function. {\rien Finally in Fig.~\ref{fig:fishmax} we show an example of output of the optimization program,
where the Fisher information at the cat-state time is maximized for different initial values of the average number of atoms in the minority component, for scattering lengths and loss rates of ${}^{87}$Rb  as in 
Fig.~\ref{fig:wigS}. As the initial atom number in the minority component is increased, first the optimal Fisher information increases, as one expects it from (\ref{eq:Fishertheta}) in the absence of losses, then it decreases due to the non-compensated, one-body losses in the minority component. By restricting to non extreme configurations, with ratios between the trap frequencies smaller than $20$, we select out the points in blue. Fig.\ref{fig:wigS} corresponds to one of the blue points with maximal Fisher information around 1500.}

\section{Adiabaticity of the interaction ramp}
\label{appen:adiab}

{\rien In {\rien the} multimode analysis of the cat-state production scheme} at zero temperature in the box $[0,L]^3$, 
after the $\pi/2$ pulse, one ramps the interaction strength $g(t)$ in each spin state
$\sigma$ from $0$ to a final positive value $g_{\rm f}$ according to the Hann semi-window (\ref{eq:Hann}). We determine here,
in the Bogoliubov approximation, the number of quasi-particles created by the ramp. Requiring that this number is
$\ll 1$ ensures adiabaticity of the process and its compatibility with cat-state production.

For a general time dependence of the coupling amplitude, the expansion (\ref{eq:devCD}) of the noncondensed field
in spin state $\sigma$ takes the form \cite{CastinDum}
\begin{multline}
\left(\begin{array}{c}\hat{\Lambda}_\sigma(\rr,t) \\ \hat{\Lambda}^\dagger_\sigma(\rr,t)\end{array}\right)
=
\sum_{\kk\neq\OO} \Big[ \hat{b}_{\sigma,\kk} (0^+) 
\left(\begin{array}{c}\mathcal{U}_k(t) \\ \mathcal{V}_k(t)\end{array} \right) \frac{e^{i\kk\cdot\rr}}{V^{1/2}}  \\
+\hat{b}^\dagger_{\sigma,\kk} (0^+) 
\left(\begin{array}{c} \mathcal{V}^*_k(t) \\ \mathcal{U}^*_k(t)\end{array} \right) \frac{e^{-i\kk\cdot\rr}}{V^{1/2}}\Big]
\end{multline}
where the complex Bogoliubov modal amplitudes obey the equations of motion
\begin{multline}
i\hbar \frac{d}{dt} \left(\begin{array}{c}\mathcal{U}_k(t) \\ \mathcal{V}_k(t)\end{array} \right) = \\
\left(\begin{array}{cc} E_k+\rho_\sigma g(t) & \rho_\sigma g(t) \\
-\rho_\sigma g(t) & -(E_k+\rho_\sigma g(t)) \end{array}\right)
\left(\begin{array}{c}\mathcal{U}_k(t) \\ \mathcal{V}_k(t)\end{array} \right)
\end{multline}
with the ideal gas initial conditions {\rien $\mathcal{U}_k(0^+)=1, \mathcal{V}_k(0^+)=0$}. Here $\rho_\sigma=N_\sigma/V$ is the density in component $\sigma$.
In the quasi-adiabatic regime it is convenient to project ($\mathcal{U}_k(t),\mathcal{V}_k(t))$ onto the instantaneous
stationary Bogoliubov mode real amplitudes $(U_k(t),V_k(t))$ of (\ref{eq:modinst}) of energy $\epsilon_k(t)$ and
on the corresponding mode $(V_k(t),U_k(t))$ of energy $-\epsilon_k(t)$:
\be
\left(\begin{array}{c}\mathcal{U}_k(t) \\ \mathcal{V}_k(t)\end{array} \right) = A_k(t) \left(\begin{array}{c}U_k(t) \\ V_k(t)\end{array} \right)
+B_k(t) \left(\begin{array}{c}V_k(t) \\ U_k(t)\end{array} \right)
\ee
with
\bea
A_k(t) &=& \phantom{+}U_k(t) \mathcal{U}_k(t) - V_k(t) \mathcal{V}_k(t) \\
B_k(t) &=& -V_k(t) \mathcal{U}_k(t) + U_k(t) \mathcal{V}_k(t) 
\eea
leading to the differential system
\be
i\hbar \frac{d}{dt} \left(\begin{array}{c}A_k(t) \\ B_k(t)\end{array} \right)=
\left(\begin{array}{cc} \epsilon_k(t) & -i\hbar\Omega_k(t) \\
-i\hbar\Omega_k(t)  & -\epsilon_k(t) \end{array}\right)
\left(\begin{array}{c}A_k(t) \\ B_k(t)\end{array} \right)
\ee
with the initial conditions $A_k(0^+)=1, B_k(0^+)=0$. The symplectic symmetry imposes 
$|\mathcal{U}_k(t)|^2-|\mathcal{V}_k(t)|^2=|A_k(t)|^2-|B_k(t)|^2=1$.
The Rabi angular frequency
\begin{multline}
\Omega_k(t)=U_k(t)\frac{d}{dt} V_k(t)-V_k(t)\frac{d}{dt} U_k(t) \\ =\frac{\frac{d}{dt}[U_k(t)+V_k(t)]}{U_k(t)+V_k(t)}
=-\frac{1}{2} \frac{\rho_\sigma \frac{d}{dt}g(t)}{E_k+2\rho_\sigma g(t)}
\end{multline}
constitutes the nonadiabatic coupling. 
The number of quasi-particle excitations in the stationary Bogoliubov mode $\kk$ present at the end of the ramp is given by
\be
n_k^{\rm exc}=|B_k(t>t_{\rm ramp})|^2 \,.
\ee

The evolution enters the adiabatic regime when the Rabi coupling is much weaker than the Bohr frequency:
\be
{\rien\hbar}|\Omega_k(t)| \ll 2 \epsilon_k(t) \,.
\label{eq:condadiab}
\ee
This is most stringent at the minimal wavenumber $k=2\pi/L$. {\rien For $2\rho_\sigma g_{\rm f}\gg \Delta=\frac{\hbar^2(2\pi/L)^2}{2m}$, this is then most stringent}
at times $\ll t_{\rm ramp}$, where the Hann expression (\ref{eq:Hann}) can be quadratised (see endnote \cite{endnote64}).
One finally gets from the adiabaticity condition (\ref{eq:condadiab}):
\be
t_{\rm ramp} \gg t_{\rm ramp}^{\rm adiab} = \frac{L^3}{\hbar^2} \frac{(\rho_\sigma g_{\rm f})^{1/2} m^{3/2}}{24\pi^2\sqrt{3}} \,.
\label{eq:condadiabtramp}
\ee
The corresponding time scale is much shorter than the cat-state formation time $t_{\rm cat}\simeq \pi/(2\chi)$ since the gas is weakly
interacting:
\be
\chi t_{\rm ramp}^{\rm adiab} \simeq \frac{(\rho_\sigma a_{\rm f}^3)^{1/2}}{3(3\pi)^{1/2}} \ll 1
\ee
with $\chi\simeq g_{\rm f}/(\hbar L^3)$ and $g_{\rm f}=4\pi\hbar^2 a_{\rm f}/m$.

In the quasi-adiabatic regime, one can treat the Rabi coupling to first order in time dependent perturbation theory, replacing in the equation
for $B_k(t)$ the amplitude $A_k(t)$ by its zeroth-order, adiabatic expression $\exp[-i\int_0^t d\tau \epsilon_k(\tau)/\hbar]$. This
gives
\be
n_k^{\rm exc} \simeq \left|\int_0^{t_{\rm ramp}} \!\! dt \,\Omega_k(t) e^{-2i \int_0^t d\tau \epsilon_k(\tau)/\hbar}\right|^2 \,.
\label{eq:nkexc}
\ee
As the Hann ramp (\ref{eq:Hann}) leads to vanishing derivatives $\frac{d}{dt}g$ at $t=0$ and $t=t_{\rm ramp}$, 
the number of excitations drops rapidly with $k$:
\be
n_k^{\rm exc} \underset{k\to+\infty}{\sim}  \left[\frac{\rho_\sigma g_{\rm f}}{8E_k^3}\left(\frac{\pi\hbar}{t_{\rm ramp}}\right)^2\right]^2 
\cos^2[(E_k+\rho_\sigma g_{\rm f}/2)t_{\rm ramp}] \,.
\ee
A numerical calculation of (\ref{eq:nkexc}) for the parameters of Fig.~\ref{fig:g1T} 
($N_\sigma=N/2=150$ and {\rien $4\pi a_{\rm f}/L=0.0667$}) confirms the condition (\ref{eq:condadiabtramp}): 
for $t_{\rm ramp}=t_{\rm ramp}^{\rm adiab}$, the total number of created excitations in each spin component is $\simeq 0.5$; 
it {\rien drops to} $\simeq 0.01$ for $t_{\rm ramp}=20\, t_{\rm ramp}^{\rm adiab}$.

\section{Details on the calculation of the spin fidelity}
\label{appen:fidelspin}

{\rien
In this appendix we derive expression (\ref{eq:fidelspinunereal}) for the spin fidelity of the state $|\psi(t)\rangle$ in equation (\ref{eq:psiunereal}) 
with respect to the cat state (\ref{eq:chatdespin}). To this end, it suffices to calculate the matrix element of a purely spinorial physical
observable $\hat{O}_{\rm spin}$
between Fock states with occupation numbers $(n_{\sigma,\kk})_\kk$ and $(n'_{\sigma,\kk})_\kk$:
\be
X=\langle a:(n'_{a,\kk})_\kk, b:(n'_{b,\kk})_\kk| \hat{O}_{\rm spin} |a:(n_{a,\kk})_\kk,b:(n_{b,\kk})_\kk\rangle
\ee
This is conveniently evaluated in the first quantization formalism, where the Fock state reads
\begin{multline}
|a:(n_{a,\kk})_\kk,b:(n_{b,\kk})_\kk\rangle=\left(\frac{N!}{\prod_{j=1}^{s} n_{a,\kk_j}! n_{b,\kk_j}!}\right)^{1/2} \\
\times \hat{S} |a,\kk_1\rangle^{\otimes n_{a,\kk_1}} |b,\kk_1\rangle^{\otimes n_{b,\kk_1}} 
\ldots |a,\kk_s\rangle^{\otimes n_{a,\kk_s}} |b,\kk_s\rangle^{\otimes n_{b,\kk_s}}
\end{multline}
where we labeled the populated wave vectors as $\kk_1, \ldots, \kk_s$ and used the notation $|u\rangle^{\otimes n}=|u\rangle {\rien \otimes \ldots \otimes}
|u\rangle$ ($n$ factors). We have introduced the symmetrisation operator 
\be
\hat{S}=\frac{1}{N!}\sum_{\sigma\in S_N} \hat{P}_\sigma
\ee
where the sum runs over all permutations $\sigma$ of $N$ elements and $\hat{P}_\sigma$ is the usual permutation operator representing $\sigma$ in the Hilbert space.
As $\hat{O}_{\rm spin}$ commutes with $\hat{S}$, due to the indistinguishability of the particles, and as $\hat{S}^2=\hat{S}$, it is enough to symmetrise
the ket only, which gives
\begin{widetext}
\be
X=\sum_{\sigma\in S_N} 
\frac{{}^{n'_{a,\kk_1}\otimes}\langle a,\kk_1| {}^{n'_{b,\kk_1}\otimes}\langle b,\kk_1| \ldots 
{}^{n'_{a,\kk_s}\otimes}\langle a,\kk_s| {}^{n'_{b,\kk_s}\otimes}\langle b,\kk_s|
\hat{O}_{\rm spin} \hat{P}_\sigma 
|a,\kk_1\rangle^{\otimes n_{a,\kk_1}} |b,\kk_1\rangle^{\otimes n_{b,\kk_1}}\ldots
|a,\kk_s\rangle^{\otimes n_{a,\kk_s}} |b,\kk_s\rangle^{\otimes n_{b,\kk_s}} 
}
{\left(\prod_{j=1}^{s} n_{a,\kk_j}! n_{b,\kk_j}! n'_{a,\kk_j}! n'_{b,\kk_j}!\right)^{1/2}}
\ee
\end{widetext}
In the matrix element of the numerator, one can move the orbital part $\langle \kk_j|$ of the bras through $\hat{O}_{\rm spin}$ to contract them with the orbital part
of the kets.
As the quantum state (\ref{eq:psiunereal}) results from the $a$-$b$ partition of an initial Fock state in internal state $a$ with occupation numbers 
$(n_\kk)_\kk$, one has $n_{a,\kk}+n_{b,\kk}=n'_{a,\kk}+n'_{b,\kk}=n_\kk, \forall\,\kk$. As a consequence, the only permutations $\sigma$ that can give a nonzero
contribution are those who leave stable (or setwise invariant) the subsets corresponding to a given $\kk$, that is
$\{1,\ldots, n_{\kk_1}\},$ $\{1+n_{\kk_1},\ldots,n_{\kk_2}+n_{\kk_1}\}$, $\ldots$, $\{1+N-n_{\kk_s},\ldots, N\}$. This gives the purely spinorial expression:
\begin{widetext}
\be
X=\frac{\left({}^{n'_{a,\kk_1}\otimes}\langle a|{}^{n'_{b,\kk_1}\otimes}\langle b|\right)\ldots
\left({}^{n'_{a,\kk_s}\otimes}\langle a| {}^{n'_{b,\kk_s}\otimes}\langle b|\right) 
\hat{O}_{\rm spin} 
\left(\sum_{\sigma_1} \hat{P}_{\sigma_1} |a\rangle^{\otimes n_{a,\kk_1}} |b\rangle^{\otimes n_{b,\kk_1}}\right)
\ldots 
\left(\sum_{\sigma_s} \hat{P}_{\sigma_s} |a\rangle^{\otimes n_{a,\kk_s}} |b\rangle^{\otimes n_{b,\kk_s}}\right)}
{\left(\prod_{j=1}^{s} n_{a,\kk_j}! n_{b,\kk_j}! n'_{a,\kk_j}! n'_{b,\kk_j}!\right)^{1/2}}
\label{eq:traced}
\ee
\end{widetext}
where in the sums the permutation $\sigma_j$ runs over $S_{n_{\kk_j}}$ the permutation group of $n_{\kk_j}$ elements. 
It remains to take for $\hat{O}_{\rm spin}$ 
the orthogonal projector on the spin cat state (\ref{eq:chatdespin}), $\hat{O}_{\rm spin}=|\mbox{spin cat}\rangle_N {}_N\langle\mbox{spin cat}|$,
to obtain (see endnote \cite{endnote65}):
\be
X= \frac{\frac{[1+i(-1)^{N'_b}]}{\sqrt{2}}\frac{[1-i(-1)^{N_b}]}{\sqrt{2}} 
\prod_{\kk}  n_\kk!}{2^N \left(\prod_\kk n_{a,\kk}! n_{b,\kk}! n'_{a,\kk}! n'_{b,\kk}!\right)^{1/2}}
\ee
where $N_\sigma=\sum_\kk n_{\sigma,\kk}$.
Since this is factorisable in a function of the $(n_{\sigma,\kk})_{\kk}$ times a function of the $(n'_{\sigma,\kk})_{\kk}$, it finally leads
to the desired expression (\ref{eq:fidelspinunereal}) of the spin fidelity of the single realisation (\ref{eq:psiunereal}),
knowing that $(-1)^{N_b}=(-1)^{S_z}$ for $N/2$ even integer.
}

{\rien In the remaining part of this appendix, we give a justification to the writing (\ref{eq:suggestive}) of the spin state vector,
which led to an enlightening interpretation of expression (\ref{eq:lumiere}) for the single realization peak fidelity in the Bogoliubov approximation. To this aim we rewrite equation (\ref{eq:traced}), where the orbital degrees of freedom have been traced out, in the form 
\be
X = \langle \chi' |\hat{O}_{\rm spin} | \chi \rangle
\label{eq:moyspin}
\ee
where we introduced the spin state vectors 
\begin{multline}
 | \chi \rangle = \left(\prod_{j=1}^s  \frac{ n_{\kk_j} !}{ n_{a,{\kk_j}}! n_{b,{\kk_j}}!}\right)^{1/2}  \\
 \times \hat{S}_{\rm partial}|a\rangle^{\otimes n_{a,\kk_1}} |b\rangle^{\otimes n_{b,\kk_1}}\ldots
 |a\rangle^{\otimes n_{a,\kk_s}} |b\rangle^{\otimes n_{b,\kk_s}} \,
\end{multline}
and $| \chi' \rangle$ defined in the same way with $(n_{\sigma,\kk})_\kk$ replaced by $(n'_{\sigma,\kk})_\kk$.
The projector $\hat{S}_{\rm partial}$ performs a partial symmetrization restricted to the aforementioned permutations, forming a subgroup $G$ of $S_N$,
that leave setwise invariant the subsets corresponding to a given $\kk$, that is
$\{1,\ldots, n_{\kk_1}\},$ $\{1+n_{\kk_1},\ldots,n_{\kk_2}+n_{\kk_1}\}$, $\ldots$, $\{1+N-n_{\kk_s},\ldots, N\}$:
\be
\hat{S}_{\rm partial} = \frac{1}{\prod_\kk n_\kk !}\sum_{\sigma \in G} \hat{P}_\sigma \,.
\ee 
Since $\prod_\kk n_\kk !$ is the cardinality of $G=\{\sigma_1 \circ \ldots \circ \sigma_s\}$, one has indeed $\hat{S}_{\rm partial}^2=\hat{S}_{\rm partial}$. Also $\hat{S}_{\rm partial}$ commutes with $\hat{O}_{\rm spin} $. Using expression (\ref{eq:psiunereal}) for the state wave vector $|\psi(t)\rangle$ in a single realization and (\ref{eq:moyspin}), we obtain
\be
\langle \psi(t) |\hat{O}_{\rm spin} |\psi(t)\rangle = \langle {\psi}_{\rm spin}(t) |\hat{O}_{\rm spin} |{\psi}_{\rm spin}(t)\rangle
\ee 
with the vector $|{\psi}_{\rm spin}(t)\rangle$ defined as follows:
\begin{widetext}
\be
|{\psi}_{\rm spin}(t)\rangle = 2^{N/2} \sum_{(n_{a,\kk})_\kk} \left[\prod_\kk P_{n_\kk}(n_{a,\kk})\right]  \\
 e^{-i\int_0^t d\tau E((n_{a,\kk})_\kk,(n_{b,\kk})_\kk,\tau)/\hbar} 
 \hat{S}_{\rm partial} |a\rangle^{\otimes n_{a,\kk_1}} |b\rangle^{\otimes n_{b,\kk_1}}\ldots
 |a\rangle^{\otimes n_{a,\kk_s}} |b\rangle^{\otimes n_{b,\kk_s}} \,,
\label{eq:psiunereal_bis}
\ee
\end{widetext}
where $P_n(n_a)=\frac{2^{-n} n!}{n_a! n_b!}$ (with $n_b=n-n_a$) is the binomial probability distribution.
Note that $|{\psi}_{\rm spin}(t)\rangle$ is not bosonic as it is only partially symmetrized. However, if we are interested in the spin dynamics in the phase space bosonic sector, which is enough to study the spin cat-state formation, we can perform the full symmetrization and consider $\hat{S}|{\psi}_{\rm spin}(t)\rangle$ which amounts to replacing $\hat{S}_{\rm partial}$ with $\hat{S}$.
In the spirit of the Bogoliubov approximation, we further perform the substitution (\ref{eq:substitution}) and quadratize the energy around $\bar{N}_a$ and $\bar{N}_b$ as in equation (\ref{eq:Hquad}) to finally obtain
\begin{widetext}
\be
\hat{S}|{\psi}_{\rm spin}^{\rm Bogol}(t)\rangle = e^{-i C(t)} \langle \sum_{N_a=0}^N \left[P_N(N_a)\right]^{1/2}\, e^{-i A(t)S_z^2} \, e^{-i \sum_{\kk\neq\OO} \gamma_k(t) (n_{a,\kk}-n_{b,\kk})S_z}
 |N_a : a , N_b : b\rangle \;\rangle_{\rm partition}
\label{eq:dephasant}
\ee
\end{widetext}
where $|N_a : a , N_b : b\rangle$ is a spin Fock state, $C(t)$ is the partition-independent integral contribution on the right-hand side of equation (\ref{eq:Hquad}), and the brackets indicate the average over the partition noise in the noncondensed modes. 
The value of $\hat{S}|{\psi}_{\rm spin}^{\rm Bogol}(t)\rangle$ at the cat-state time (such that $A(t)=\pi/2$) reproduces equation
(\ref{eq:suggestive}), and its scalar product with the target state (\ref{eq:cat0}) reproduces the form
(\ref{eq:lumiere}). The writing (\ref{eq:dephasant}) of the state shows that the effect of finite temperature is captured by a two-mode model, see equation (\ref{eq:baleze}), supplemented by a dephasing environment. This kind of model was already used in the context of spin squeezing \cite{Ferrini2011,Frontiers} in particular to predict the optimum spin squeezing at finite temperature \cite{Frontiers}. 
Contrarily to references \cite{Ferrini2011,Frontiers}, the stochastic element enters here in two different ways: the average over the partition noise in the noncondensed modes results in a coherent superposition of kets, while the average over the initial thermal excitations in component $a$ is a classical average at the level of the density matrix which results in a statistical mixture.

To be complete, we give the expression of the mean value of the total spin, which is along the $x$ axis for the considered initial state of the system. To this end, we give another writing of equation (\ref{eq:psiunereal_bis}):
\begin{widetext}
\be
|{\psi}_{\rm spin}(t)\rangle = \sum_{(n_{a,\kk})_\kk} \left[\prod_\kk P_{n_\kk}(n_{a,\kk})\right]^{1/2}  
 e^{-i\int_0^t d\tau E((n_{a,\kk})_\kk,(n_{b,\kk})_\kk,\tau)/\hbar} 
|n_{a,\kk_1} : a, n_{b,\kk_1} :b \rangle \otimes \ldots \otimes |n_{a,\kk_s} : a, n_{b,\kk_s} :b \rangle  
\label{eq:psiunereal_ter}
\ee
\end{widetext}
where $|n_{a,\kk} : a, n_{b,\kk} :b \rangle$ are spin Fock states. We calculate the action of 
$\hat{S}_+=\hat{S}_x+i\hat{S}_y=\sum_{i=1}^N |a\rangle_i \,{}_i\langle b|$ on equation (\ref{eq:psiunereal_ter}). The first $n_{\kk_1}$ terms
of $\hat{S}_+$ act on the first Fock state $|n_{a,\kk_1} : a, n_{b,\kk_1} :b \rangle$ and give 
$[(1+n_{a,\kk_1}) n_{b,\kk_1} ]^{1/2}|n_{a,\kk_1}+1 : a, n_{b,\kk_1}-1 :b \rangle$, and so forth for the $n_{\kk_2},n_{\kk_3},\ldots$ subsequent terms. By performing the energy quadratization (\ref{eq:Hquad}) but not the Bogoliubov substitution (\ref{eq:substitution}), we obtain the single realization result as a sum of the contributions of the various single particle modes:
\be
\langle \psi(t) |\hat{S}_+|\psi(t) \rangle = \sum_\qq {\cal C}_\qq(t) \,.
\ee
The condensate contribution is
\be
{\cal C}_{\mathbf{0}}(t)=\frac{n_\mathbf{0}}{2} \cos^{N-1} [A(t)] \prod_{\kk \neq \mathbf{0}} 
\left( \frac{\cos[\gamma_k(t)+A(t)]}{\cos A(t)} \right)^{n_\kk} 
\label{eq:C0}
\ee 
while the noncondensed $\qq\neq \mathbf{0}$ mode contribution is
\begin{multline}
{\cal C}_\qq(t)=\frac{n_\qq}{2}  \cos^{N-1} [A(t)+\gamma_q(t)]  \\
\times \prod_{\kk \neq \mathbf{0}} 
\left( \frac{\cos[\gamma_k(t) + \gamma_q(t)+A(t)]}{\cos [\gamma_q(t)+A(t)]} \right)^{n_\kk-\delta_{\kk,\qq}} 
\label{eq:Cq}
\end{multline}
where $\delta_{\kk,\qq}$ is a Kronecker delta.
If in equation (\ref{eq:C0}) one approximates $n_\mathbf{0}$ by its mean value and one performs 
the thermal average over the $n_\kk$,
one recovers exactly the result (\ref{eq:g1bogfin}) for the condensate first order coherence function. Interestingly the contributions 
of the noncondensed modes to the mean spin have different revival times than the condensate. As a consequence they do not contribute to the major peaks in $\langle \hat{S}_x\rangle(t)$, they contribute to side peaks of very small relative amplitudes
$O(f_{\rm nc})$ ($f_{\rm nc}$ is the initial noncondensed fraction).
}

%

\begin{thebibliography}{99}
\bibitem{endnote51}
In reference  \cite{Simon}, the existence of a spin cat state is deduced from the existence of a revival peak of the Husimi function. The revival shown (see Fig. 3 in that reference, lower right panel) is however significantly reduced with respect to its maximal decoherence-free $1/\pi$ value, indicating a low fidelity of the corresponding cat state.
\bibitem{endnote52}
The contribution of two-body losses, whose origin is spin changing collisions, can be made negligible by choosing extremum Zeeman sub-levels.
\bibitem{endnote53}
Extremely long lifetimes of the order of hours might be obtained in cryogenic environment \cite{Libbrecht,antiH}. As one can estimate from equations (\ref{eq:lambda}) and (\ref{eq:simple_fidelity}), the influence of the corresponding one-body losses at $t_{\rm cat}=0.128\,$s  would then be negligible.
\bibitem{endnote54}
We do not include crossed three-body processes as $aab$, etc., because their contribution to the total loss rate is very small far from a Feshbach resonance.
\bibitem{endnote55}
The contribution of the noncondensed modes to the full first order coherence of the field, that is to the mean collective spin, is discussed in Appendix \ref{appen:fidelspin}.
\bibitem{endnote56}
We have replaced the operator multiplying $\hat{N}_a-\hat{N}_b$ by its mean value, which introduces an error
$O(N^{-1}t f_{\rm nc})$ where $f_{\rm nc}$ is the noncondensed fraction; at the revival time this introduces a {\rien small} $O(f_{\rm nc})$ error.
\bibitem{endnote57}
The nonzero temperature correction to $\chi$ was missed in reference \cite{Casagrande}.
\bibitem{endnote58}
The number of particles in the condensate modes is no longer well defined,
but what matters in the number conserving Bogoliubov theory is the total number of particles $N_\sigma$ in each spin state, which
is well defined.
\bibitem{endnote59}
The result (\ref{eq:fidelspinunereal}) holds under the assumption that the single-particle orbital states
are the same in the internal states $a$ and $b$. In practice, this means that $a$ and $b$ experience the same trapping potential. If the $a$ and $b$ traps
were spatially translated to ensure an effective $g_{ab}=0$ coupling constant, they must be translated back to the same location at the cat-state time.
\bibitem{endnote60}
At time $t=0^+$, for a single realisation, one has $\frac{\mathrm{Cov}\,(N_a-N_b,N_{a,\mathrm{nc}}-N_{b,\mathrm{nc}})}
{[\mathrm{Var}\,(N_a-N_b)\mathrm{Var}\,(N_{a,\mathrm{nc}}-N_{b,\mathrm{nc}})]^{1/2}}=\left(\frac{\sum_{\kk\neq\mathbf{0}} n_\kk}{N}\right)^{1/2}$
where $\mathrm{Var}$ stands for the variance and $\mathrm{Cov}$ for the covariance.
\bibitem{endnote61}
This state has a norm less than one because it corresponds to the restriction of the spin density operator to the bosonic sector.
This spin density operator, obtained as a trace of the full density operator over the orbital variables,
can indeed populate various irreducible representations of the permutation group $S_N$, not simply the bosonic one, when the orbital variables do
not occupy a single mode of the field. We give more details in Appendix \ref{appen:fidelspin}.
\bibitem{endnote62}
This reasoning fails if $\gamma_k=2\pi$ for the considered single excitation. This is not possible at the cat-state time such that $A(t_{\rm cat})=\pi/2$, as in the weakly interacting regime and for $t_{\rm cat}\gg t_{\rm ramp}$, $\gamma_k/A\simeq [E_k/(E_k+2g_{\rm f} \rho_{\sigma})]^{1/2}<1$. This would be possible for the third cat-state time
$A(t_{\rm cat}^{(3)})=5\pi/2$ for well chosen parameters such that $\gamma_k/A\simeq 4/5$. For the geometry and interaction parameters of Fig.~\ref{fig:fidel}, this is approximately realized in the first excited manifold $k=2\pi/L$ for a total particle number $N=168$ (if one restricts to $N=0$ modulo $4$). As the first excited manifold then does not contribute to the dephasing, the spin fidelity will be limited by the second excited manifold at energy $2\Delta$. This clever tuning allows to increase the required temperature by a factor of two.
\bibitem{endnote63}
The last expression in equation (\ref{eq:minorant}) results from the average of $ [P_{N_k}(N_k/2)]^2$ over the $N_k$ with their probability distribution $\pi_k(N_k)=(1-q_k)^{d_k}q_k^{N_k}\frac{(N_k+d_k-1)!}{N_k! (d_k-1)!}$ where $q_k=\exp(-\beta E_k)$. This results from the fact that $N_k$ is the sum of $d_k$ independent random variables with the same law $n \mapsto (1-q_k)q_k^n$, the binomial coefficient giving the number of ways to arrange $N_k$ identical objects in $d_k$ boxes.
\bibitem{endnote64}
In the opposite limit $2\rho_\sigma g_{\rm f} \ll \Delta$, the adiabaticity condition is most stringent at time $t=t_{\rm ramp}/2$, where
the derivative of $g(t)$ is maximal. One then gets the condition $t_{\rm ramp}\gg t_{\rm ramp}^{\rm adiab}=\frac{\pi}{8} \frac{\hbar\rho_\sigma g_{\rm f}}{\Delta^2}$. For the parameters
of Fig.~\ref{fig:g1T}, one is in the intermediate regime $2\rho_\sigma g_{\rm f}\simeq \Delta$ but the adiabaticity times $t_{\rm ramp}^{\rm adiab}$
issued from the two limiting cases
agree within $30 \% $.
\bibitem{endnote65}
One may use the property that the scalar product of $|u\rangle^{\otimes N}$ with {\rien a tensorial-product} ket composed of $N_a$ 
factors equal to $|a\rangle$ and $N_b$ factors equal to $|b\rangle$ is $(\langle u|a\rangle)^{N_a} (\langle u|b\rangle)^{N_b}$ irrespective of the order of these factors.
\end{thebibliography}
\end{document}